\pdfoutput=1
\RequirePackage{ifpdf}
\ifpdf 
\documentclass[pdftex]{sigma}
\else
\documentclass{sigma}
\fi

\usepackage{tikz,longtable}
\usepackage{caption}
\usepackage{subcaption}
\usetikzlibrary{trees,shapes,shapes.symbols,matrix}
\tikzset{
 treenode/.style = {align=center,rounded corners=1mm, text centered, font=\sffamily, anchor = north},
 geo/.style = {treenode, rectangle, draw=blue},
 sph/.style = {treenode, rectangle, draw=red}
}

\numberwithin{equation}{section}

\newtheorem{Theorem}{Theorem}[section]

\newtheorem{Proposition}[Theorem]{Proposition}
 { \theoremstyle{definition}
\newtheorem{Definition}[Theorem]{Definition}

\newtheorem{Example}[Theorem]{Example}
\newtheorem{Remark}[Theorem]{Remark} }

\begin{document}

\renewcommand{\thefootnote}{$\star$}

\newcommand{\arXivNumber}{1607.00712}

\renewcommand{\PaperNumber}{117}

\FirstPageHeading

\ShortArticleName{Separation of the Hamilton--Jacobi Equation on Spaces of Constant Curvature}

\ArticleName{Orthogonal Separation of the Hamilton--Jacobi\\ Equation on Spaces of Constant Curvature\footnote{This paper is a~contribution to the Special Issue on Analytical Mechanics and Dif\/ferential Geometry in honour of Sergio Benenti.
The full collection is available at \href{http://www.emis.de/journals/SIGMA/Benenti.html}{http://www.emis.de/journals/SIGMA/Benenti.html}}}

\Author{Krishan RAJARATNAM~$^\dag$, Raymond G.~MCLENAGHAN~$^\ddag$ and Carlos VALERO~$^\ddag$}

\AuthorNameForHeading{K.~Rajaratnam, R.G.~McLenaghan and C.~Valero}

\Address{$^\dag$~Department of Mathematics, University of Toronto, Canada}
\EmailD{\href{kr.rajaratnam@mail.utoronto.ca}{kr.rajaratnam@mail.utoronto.ca}}

\Address{$^\ddag$~Department of Applied Mathematics, University of Waterloo, Canada}
\EmailD{\href{mailto:rgmclenaghan@uwaterloo.ca}{rgmclenaghan@uwaterloo.ca}, \href{mailto:cjvalero@uwaterloo.ca}{cjvalero@uwaterloo.ca}}

\ArticleDates{Received September 30, 2015, in f\/inal form December 12, 2016; Published online December 21, 2016}

\Abstract{We review the theory of orthogonal separation of variables of the Hamilton--Jacobi equation on spaces of constant curvature, highlighting key contributions to the theo\-ry by Benenti. This theory revolves around a special type of conformal Killing tensor, hereafter called a concircular tensor. First, we show how to extend original results given by Benenti to intrinsically characterize all (orthogonal) separable coordinates in spaces of constant curvature using concircular tensors. This results in the construction of a special class of separable coordinates known as Kalnins--Eisenhart--Miller coordinates. Then we present the Benenti--Eisenhart--Kalnins--Miller separation algorithm, which uses concircular tensors to intrinsically search for Kalnins--Eisenhart--Miller coordinates which separate a~given na\-tural Hamilton--Jacobi equation. As a new application of the theory, we show how to obtain the separable coordinate systems in the two dimensional spaces of constant curvature, Minkowski and (Anti-)de Sitter space. We also apply the Benenti--Eisenhart--Kalnins--Miller separation algorithm to study the separability of the three dimensional Calogero--Moser and Morosi--Tondo systems.}


\Classification{53C15; 70H20; 53A60}

\renewcommand{\thefootnote}{\arabic{footnote}}
\setcounter{footnote}{0}

\section{Introduction}

Separation of variables of the Hamilton--Jacobi equation is an old but still powerful tool for obtaining exact solutions. Until recently, it was not known how to exploit this method to its maximum potential.

In this article, we review important advances made to the theory of separation of variables in spaces of constant curvature f\/irst presented in the articles \cite{Rajaratnam2014d, Rajaratnam2014a}. We will also point out how these contributions build on, and unify, work done by Sergio Benenti and his co-workers Ernie Kalnins, Willard Miller and Micheal Crampin. Finally, we will present new results by showing how the theory presented in~\cite{Rajaratnam2014a} can be used to enumerate the separable coordinate systems in $n$-dimensional Lorentzian spaces with zero curvature, $\mathbb{E}^{n}_1$, in de-Sitter, $\operatorname{dS}_n$, and anti-de Sitter, $\operatorname{AdS}_n$, spaces.

We assume the reader is familiar with the theory of separation of variables on Riemannian manifolds, which can be found in \cite{Benenti2004} for example. In the present article, we will f\/irst brief\/ly review this theory in Section~\ref{sec:intChar}, introduce concircular tensors in Section~\ref{sec:CTintro}, show how these tensors can be used to separate geodesic Hamilton--Jacobi equations in Section~\ref{sec:geo}, and then natural Hamilton--Jacobi equations in Section~\ref{sec:sepNats}. Furthermore, we enumerate the isometrically inequivalent separable coordinate systems in $\mathbb{E}_1^2$ (resp. $\operatorname{dS}_2$) in Section~\ref{sec:sepE21} (resp.\ Section~\ref{sec:sepdS2}).

In this article, $(M, g)$ denotes a pseudo-Riemannian manifold and $T^{*}M$ denotes the cotangent bundle of $M$. If $(q,p)$ denotes canonical coordinates on $T^{*}M$, then the (natural) \emph{Hamiltonian}~$H$ with potential $V \in \mathcal{F}(M)$, where $\mathcal{F}(M)$ denotes the smooth functions from $M$ into $\mathbb{R}$, is def\/ined by
\begin{gather} \label{eq:Hamilton}
H(q,p) := \frac{1}{2}g^{ij}(q)p_{i}p_{j} + V(q).
\end{gather}

The \emph{geodesic Hamiltonian} is obtained by setting $V \equiv 0$ in the above equation. The \emph{Hamilton--Jacobi equation} is a partial dif\/ferential equation def\/ined on~$M$ in terms of the Hamiltonian as follows:
\begin{gather} \label{eq:Hmailton-Jacobi}
H\left(q,\frac{\partial W}{\partial q^i}\right) := \frac{1}{2}g^{ij}(q)\frac{\partial W}{\partial q^i}\frac{\partial W}{\partial q^j} + V(q) = 0.
\end{gather}

Coordinates $(q^i)$ (for $M$) are called \emph{separable} if they are orthogonal and the Hamilton--Jacobi equation admits a complete integral of the form
\begin{gather*}
W\big(q^1,\ldots,q^n,c_1,\ldots,c_n\big) = \sum_{i=1}^{n} W_i\big(q^i,c_1,\ldots,c_n\big).
\end{gather*}

In the theory of separation of variables for the Hamilton--Jacobi equation, one wishes to solve the following \emph{fundamental problems}:
\begin{enumerate}\itemsep=0pt
	\item \label{fpro:I} Given a (pseudo-)Riemannian manifold, what are the ``inequivalent'' coordinate systems that separate the geodesic Hamiltonian?
	\item \label{fpro:II} How do we determine, intrinsically (coordinate-independently), the ``inequivalent'' coordinate systems in which a given natural Hamiltonian separates?
	\item \label{fpro:III} If we have determined that the natural Hamiltonian is separable in coordinates $(u^{1},\dots,u^{n})$, what is the transformation to these coordinates from the original position-momentum coordinates $(q^{1},\dots,q^{n})$ in which the natural Hamiltonian is def\/ined?
\end{enumerate}

In this article, we will show how concircular tensors can be used to obtain an elegant solution to these problems.

\section{The Intrinsic characterization of separation} \label{sec:intChar}

The f\/irst crucial result is due to St\"{a}ckel \cite{Stackel1893} who showed that the Hamilton--Jacobi equation of a natural Hamiltonian is orthogonally separable with respect to coordinates $(q^i)$ if\/f there exists a~$n\times n$ matrix $\tilde{S}_{ij}(q^{i})$ which determines the forms of the metric $g$ and potential $V$ in equation~\eqref{eq:Hamilton} (see also \cite[p.~9]{Kalnins1986}). He further showed that if equation~\eqref{eq:Hmailton-Jacobi} is separable in this way, then $H$ admits $n$ f\/irst integrals $F_{1},\dots,F_{n}$ (where $F_{1} := H$) expressible in terms of $\tilde{S}$, each of which has the following form in canonical coordinates $(q^i,p_j)$:
\begin{gather} \label{eq:firstInts}
F(q,p) = \frac{1}{2}K^{ij}(q)p_{i}p_{j} + U(q)
\end{gather}
with
\begin{gather*}
\{F_{i}, F_{j}\} = 0, \qquad {\rm d} F_1 \wedge \cdots \wedge {\rm d} F_{n} \neq 0,
\end{gather*}
where $\{\cdot,\cdot\}$ denotes the Poisson bracket. One can show that the involutory condition $\{F, H \} = 0$ is equivalent to the following equations on $M$ \cite[Section~4]{Benenti2004}:
\begin{subequations} \label{eq:StackImps}
	\begin{gather} \label{eq:StackImpsI}
	\nabla_{(i}K_{jk)} = 0,
\\
\label{eq:StackImpsII}
	{\rm d} U = K {\rm d} V,
	\end{gather}
\end{subequations}
where $\nabla$ is the Levi-Civita connection induced by $g$. The equation~\eqref{eq:StackImpsI} shows that~$K$ is a~Killing tensor (KT) on~$(M,g)$. Using this fact Eisenhart obtained an intrinsic characterization of orthogonal separation for the geodesic Hamiltonian in pseudo-Riemannian spaces~\cite{Eisenhart1934}. He showed that the metric $g$ is of St\"{a}ckel form if\/f the Hamiltonian admits $n-1$ quadratic f\/irst integrals of the form \eqref{eq:firstInts} such that the associated Killing tensors for each of the integrals have pointwise simple eigenfunctions, and the corresponding eigenvector f\/ields are normal. The theorem is proved by writing equation~\eqref{eq:StackImpsI} with respect to a system of coordinates adapted to the $n$ foliations orthogonal to the common eigenvector f\/ields of the Killing tensors. This procedure yields a system of partial dif\/ferential equations in the $n$ eigenfunctions called the {\em Eisenhart's equations}~(E). The integrability conditions for these equations are a system of second order partial dif\/ferential equations in the components of the metric tensor called the {\em Eisenhart integrability conditions} (EIC) which remarkably do not contain the eigenfunctions. The met\-ric~$g$ is in St\"{a}ckel form if\/f the integrability conditions are satisf\/ied.

Eisenhart's theorem has been reformulated by Benenti~\cite{Benenti1997a}. In order to state his result, the following def\/inition is required: A~\emph{characteristic Killing tensor $($ChKT$)$} is a Killing tensor which has point-wise simple eigenfunctions and the eigenvector f\/ields of which are orthogonally integrable (normal). The latter condition is equivalent to the statement that $(M,g)$ admits coordinates in which $K$ is diagonalized.

\begin{Theorem}[orthogonal separation of geodesic Hamiltonians] \label{Intthm:HJosepI}
	The geodesic Hamilto\-nian~$H$ on a space $(M,g)$ is separable in orthogonal coordinates iff there exists a ChKT which is diagonalized in these coordinates.
\end{Theorem}

The proof given by Benenti depends on a characterization of separation of variables for a~general~$H$ in terms of a system of partial dif\/ferential equations given by Levi-Civita~\cite{Levi-Civita1904}. He proved the crucial result that the separation of the geodesic Hamiltonian is a necessary condition for the separation of a natural Hamiltonian~\eqref{eq:Hamilton}.

Given a ChKT, $K$, let $\mathcal{E} = (E_1,\dots,E_n)$ denote the collection of eigenspaces of~$K$. The above theorem shows that any coordinates $(q^i)$ with the property that $\operatorname{span} \{\partial_i\} = E_i$ are separable. Hence we call the collection $\mathcal{E}$ a \emph{separable web}. More generally, any collection $\mathcal{E} = (E_1,\dots,E_n)$ of pair-wise orthogonal non-degenerate $1$-distributions which admit local coordinates $(q^i)$ satisfying $\operatorname{span} \{\partial_i\} = E_i$ is called an \emph{$($orthogonal$)$ web}. Since separable webs are uniquely determined by ChKTs, we will often work with them instead of coordinates.

The equation~\eqref{eq:StackImpsII} is a compatibility condition between the KT $K$ and potential~$V$. In~\cite{Benenti1997a} Benenti obtains an intrinsic characterization of separation for natural Hamiltonians.

\begin{Theorem}[orthogonal separation of natural Hamiltonians] \label{Intthm:HJosepII}
	A natural Hamiltonian with potential $V$ is separable in orthogonal coordinates $(q^i)$ iff there exists a ChKT $K$ diagonalized in these coordinates which satisfies
	\begin{gather*} 
	{\rm d} (K {\rm d} V) = 0.
	\end{gather*}
\end{Theorem}

The above equation is called the \emph{dKdV equation} associated with the KT $K$ and potential~$V$.

For geodesic separation, each f\/irst integral given by equation~\eqref{eq:firstInts} has a corresponding KT~$K$. It can be deduced from St\"{a}ckel's theorem that the $n$ KTs are point-wise independent on $M$ and span an $n$-dimensional vector space of KTs which are simultaneously diagonalized in the separable coordinates. This vector space of KTs is called the \emph{Killing--St\"ackel space $($KS-space$)$} associated with a separable web.

We conclude with the following observations. Firstly, Theorem~\ref{Intthm:HJosepI} implies that the problem of classifying separable coordinates for a geodesic Hamiltonian is equivalent to the problem of classifying ChKTs. Secondly, Theorem~\ref{Intthm:HJosepII} shows that the problem of classifying ChKTs is important for separating natural Hamiltonians as well. Finally, we note that the general theory of the intrinsic characterization of separation on pseudo-Riemannian manifolds can be found in~\cite{Benenti1997a}.

\section{Concircular tensors} \label{sec:CTintro}

In the previous section we gave an intrinsic characterization of separation, which allows one, in principle, to obtain all separable coordinates systems def\/ined on a given pseudo-Riemannian manifold.
In fact, the direct integration of the Eisenhart integrability conditions of\/fers the only method of obtaining separable coordinates when the space $(M,g)$ admits a low or zero dimensional isometry group. This method has also been used to obtain separable coordinates in the following spaces of constant curvature: $\mathbb{E}^3$ \cite{Eisenhart1934}, $\mathbb{S}^3$, $\mathbb{H}^3$ \cite{Olevsky1950}, $\mathbb{S}^n$, $\mathbb{E}^n$, $\mathbb{H}^n$ \cite{Kalnins1986, Kalnins1982, Kalnins1986a}, $\mathbb{E}^2_1$, $\mathbb{E}^3_1$ \cite{Horwood2008a, Kalnins1975, McLenaghan2002a},
$\mathbb{E}^{n}_{\mathbb{C}}$, $\mathbb{S}^{n}_{\mathbb{C}}$ \cite{Kalnins1984}. From the last two cases one can obtain the real cases by restriction. In the case of the spaces which admit isometry groups of maximal dimension, other more algebraic methods are possible. Indeed, Kalnins and Miller obtain all the inequivalent separable coordinate systems for the Hamilton--Jacobi equation for the geodesics and the Laplace equation on real positive def\/inite Riemannian spaces of constant curvature ($\mathbb{E}^n$, $\mathbb{S}^n$, $\mathbb{H}^n$), by solving the EIC by means of an inductive procedure such that (for example) given all separable systems for~$\mathbb{S}^k$, $k<n$, one can give the rules for the construction of all systems for~$\mathbb{S}^n$, thereby solving the f\/irst fundamental Problem~(\ref{fpro:I}). In addition, using the fact that spaces considered admit isometry groups of maximal dimension, they develop a graphical calculus based on Lie group theory which summarizes the complete solution. Their calculus has recently been reinterpreted from an algebraic geometry point of view by Sch\"{o}bel~\cite{Schobel2014}. Waksj\"o and Rauch-Wojciechowski in~\cite{Waksjo2003} used this procedure to solve the last two fundamental Problems~(\ref{fpro:II}) and (\ref{fpro:III}) for $n$-dimensional Euclidean and spherical spaces.

McLenaghan, Smirnov and collaborators have developed a theory called the invariant theory of Killing tensors for classifying characteristic Killing tensors and hence separable webs on pseudo-Euclidean spaces of low dimension \cite{Cochran2011, Horwood2005,Horwood2009, McLenaghan2002b,McLenaghan2004}. In this theory the dif\/ferent types of possible webs are classif\/ied by means of a set of functions of the coef\/f\/icients of the quadratic functions which are the components of the corresponding ChKT expressed in pseudo-Cartesian coordinates and which are invariant under the action of the isometry group~\cite{Horwood2008c}. Once the type of web has been found by the above method, the transformation from canonical separable coordinates (determined by Eisenhart's method) to pseudo-Cartesian coordinates is determined up to an isometry by the procedure described in~\cite{Horwood2007}. The theory can be extended to natural Hamiltonians on these spaces by applying Theorem~\ref{Intthm:HJosepII} to the general KT~$K$ and given potential function $V$ to obtain a restricted form of~$K$, to which the above classif\/ication procedure may be applied. This theory solves all the fundamental problems given in the introduction. However, a~generalization to higher dimensions seems problematical for the following reasons:
\begin{itemize}\itemsep=0pt
	\item It is dif\/f\/icult to obtain an algebraic expression for the general ChKT $K$ in a space of constant curvature. Indeed, one can show that the general Killing tensor, $K$, in a space of constant curvature is a sum of symmetrized products of the Killing vectors of the space~\cite{Thompson1986}. An invariant condition that $K$ have normal vector f\/ields is that it satisf\/ies the Tonolo--Schouten conditions or the equivalent Haantjes condition \cite{Haantjes1955, Nijenhuis1951} both of which are non-linear in the coef\/f\/icients of~$K$. The general solution of these equations for $\mathbb{S}^3$ is given in~\cite{Schobel2014}. However, the solution for arbitrary $n$ appears impossible. The condition that the eigenfunctions of $K$ be point-wise simple also seems intractable.
	\item A generalization of the classif\/ication of ChKTs in terms of isometry invariants for general~$n$ also appears problematical. While the invariants of $K$ may be computed~\cite{Horwood2008c}, it is unclear how one would obtain the particular combinations of invariants required for a classif\/ication scheme for the separable webs. (See~\cite{Horwood2005, Horwood2009} for the solution in $\mathbb{E}^3$ and $\mathbb{E}^3_1$.)
\end{itemize}

A careful study of the approaches described above and the works of others \cite{Benenti2005a,Crampin2003} show that concircular tensors have a fundamental role to play in the theory.

A \emph{concircular tensor} (CT), $L \in S^2(M)$ (where $S^2(M)$ is the space of symmetric contravariant $2$-tensors on $M$), is def\/ined by the following equation:
\begin{gather*}
\nabla_{k}L_{ij} = \alpha_{(i}g_{j)k}
\end{gather*}
for some covector $\alpha$. In the above equation $L$ is a covariant $2$-tensor. Throughout this article we will use the same symbol regardless of whether $L$ is covariant, contravariant or an endomorphism, where we identify these tensors by means of the canonical isomorphism induced by the metric tensor~$g$. The type of the tensor should be clear from the context.

One can obtain a general solution to the above equation in $\mathbb{E}^{n}_{\nu}$. First, def\/ine the \emph{dilatational vector field} in $\mathbb{E}^{n}_{\nu}$ in Cartesian coordinates $(x^i)$ by $r := x^i \partial_i$. Then the general solution is given in contravariant form by \cite{Benenti2005a}:
\begin{gather} \label{eq:CTGenEunn}
L = A + 2 w \odot r + m r \odot r,
\end{gather}
where $A = A^{ij} \partial_i \odot \partial_j$ with $A^{ij}$ a constant symmetric matrix, $w$ is a~constant vector, $m$ is a constant scalar and $\odot$ denotes the symmetric product. We denote the unit sphere with signed radius $r^2$ in $\mathbb{E}^{n+1}_{\nu}$ by $\mathbb{E}^{n+1}_{\nu}(\frac{1}{r^2})$. Then the restriction of the above tensor to $\mathbb{E}^{n+1}_{\nu}(\frac{1}{r^2})$ gives the general CT~\cite{Rajaratnam2014}. CTs solve the problems confronted with ChKTs listed above. Indeed, in this article, we will show that CTs can be used to solve the fundamental problems in spaces of constant curvature.

We say a CT is an \emph{orthogonal concircular tensor $($OCT$)$} if it is point-wise diagonalizable. An important property of OCTs is that they always admit local coordinates which diagonalize them. More precisely, suppose~$L$ is an OCT, then there exist local coordinates $(x^{i})$ such that $L$ has the following form (see~\cite{Benenti2005a}):
\begin{gather} \label{Inteq:CTeqn}
L = \sum\limits_{a \in M} \sigma_{a} \partial_{a} \otimes {\rm d} x^{a} + \sum\limits_{j = 1}^{k} e_{j} \sum\limits_{i \in I_j} \partial_{i} \otimes {\rm d} x^{i},
\end{gather}
where $M, I_1,\dots,I_k \subseteq \{1,\dots,n\}$, $\{1,\dots,n\} = M \sqcup (\sqcup_{j=1}^k I_j)$, the $\sigma_{a}(x^{a})$ are non-constant, and the $e_j$ are constants. Additionally, at each point, $\sigma_a \neq \sigma_b$ for $a \neq b$, $e_i \neq e_j$ for $i \neq j$ and $\sigma_a \neq e_i$.

Benenti showed that concircular tensors can be used to construct Killing tensors~\cite{Benenti1992c}. Indeed, if $L$ is a CT, it can be shown that the following sequence of tensors are KTs~\cite{Benenti2005a}:
\begin{gather}
K_{0} = G, \qquad K_1 = \operatorname{tr}(L) G - L, \label{introeq:BTLeq}\\
K_{a} = \frac{1}{a} \operatorname{tr}(K_{a-1}L) G - K_{a-1}L, \qquad 1 < a < n. \label{introeq:BTLseqb}
\end{gather}

Since the KT $K_1$ is special, we call it the \emph{Killing Bertrand--Darboux tensor (KBDT)} associated with $L$. This KT will be useful for connecting CTs with the general theory of separation given in the previous section. An important observation is that it has the same eigenspaces as $L$.

Concircular tensors were f\/irst introduced into the theory of separation of variables by Benenti~\cite{Benenti1992c}, and referred to as ``inertia tensors'', in order to calculate the Killing--St\"{a}ckel space for the elliptic and parabolic coordinates in Euclidean space. They were later studied on general Riemannian manifolds by Crampin~\cite{Crampin2005}, and then again by Benenti~\cite{Benenti2005a}. These studies considered the case where the CTs had point-wise simple eigenvalues. It was later shown by the authors in \cite{Rajaratnam2014a} that even in the non-simple case, i.e., point-wise diagonalizable, CTs could be used to separate the Hamilton--Jacobi equation.

\section{Separation of geodesic Hamiltonians} \label{sec:geo}
\subsection{Benenti tensors}
We say a CT $L$ is a \emph{Benenti tensor} if it has point-wise simple eigenvalues. A key observation made by Benenti is that any such tensor induces a separable web~\cite{Benenti1992c}. Indeed, since the KBDT is a KT with simple eigenfunctions and can be diagonalized in a coordinate system (see equation~\eqref{Inteq:CTeqn}), it's a ChKT, hence the result follows by Theorem~\ref{Intthm:HJosepI}. Furthermore, it can be shown that the KTs given by equations~\eqref{introeq:BTLeq} and~\eqref{introeq:BTLseqb} form a basis for the KS-space associated with this separable web~\cite{Benenti2005a}.

An important class of Benenti tensors are the \emph{irreducible concircular tensors (ICTs)}. A CT $L$ is called irreducible if it's a Benenti tensor and its eigenfunctions are functionally independent. By equation~\eqref{Inteq:CTeqn} any Benenti tensor with non-constant eigenfunctions is irreducible. This class of CTs is of interest, because in this case, by equation~\eqref{Inteq:CTeqn}, the eigenfunctions can be used as separable coordinates! We call these the \emph{canonical coordinates} associated with the ICT. We will see shortly that ICTs can be used as building blocks to construct more general classes of separable coordinates. The following is the prototypical example of an ICT:

\begin{Example}[elliptic coordinates in $\mathbb{E}^2$] \label{ex:ellipCoord}
	Let $M = \mathbb{E}^2$ and f\/ix an orthonormal basis $\{d,e\}$ for this Euclidean space. Let $(x,y)$ be Cartesian coordinates for $\mathbb{E}^2$ so that $d = \partial_x$ and $e = \partial_y$. Then consider the following CT:
	\begin{gather}
	L = \lambda_1 d \odot d + \lambda_2 e \odot e + r \odot r.
	\end{gather}
	
	Without loss of generality we can assume $\lambda_1 < \lambda_2$. We will show how to obtain the transformation from separable to Cartesian coordinates after showing that $L$ is a Benenti tensor. The characteristic polynomial of $L$ is given as follows:
	\begin{gather*}
	p(z) = \det (z I - L) = (z- \lambda_1)(z- \lambda_2) - x^2 (z - \lambda_2) - y^2 (z - \lambda_1).
	\end{gather*}
	
	From the above equation, we note the following:
	\begin{gather}
	p(\lambda_1) = x^2 (\lambda_2 - \lambda_1),\qquad p(\lambda_2) = y^2 (\lambda_1 - \lambda_2). \label{introeq:polVal}
	\end{gather}
	
	Now, assume that $x,y \neq 0$. Then we observe that $p(\lambda_1) > 0$, $p(\lambda_2) < 0$ and $\lim\limits_{z \rightarrow \infty} p(z) = \infty$. Hence by the intermediate value theorem, at each point, $p(z)$ has two distinct roots $u^1 < u^2$ satisfying:
		\begin{gather*}
	\lambda_1 < u^1 < \lambda_2 < u^2.
	\end{gather*}
	
	Thus $L$ is a Benenti tensor. Since ${\rm d} p \neq 0$, it follows that $L$ cannot have constant eigenfunctions \cite[Section~9.4]{Rajaratnam2014}, thus from the preceding discussion we see that $L$ is an ICT. Now observe that we can write $p(z) = (z - u^1)(z - u^2)$. Then equation~\eqref{introeq:polVal} can be used to obtain the transformation from the separable coordinates $(u^1 , u^2)$ to Cartesian coordinates $(x,y)$:
	\begin{gather*}
	x^2 = \frac{\big(\lambda_1 - u^1\big)\big(\lambda_1 - u^2\big)}{(\lambda_2 - \lambda_1)}, \qquad y^2 = \frac{\big(\lambda_2 - u^1\big)\big(\lambda_2 - u^2\big)}{(\lambda_1 - \lambda_2)}.
	\end{gather*}
\end{Example}

The above example can be generalized to higher dimensions and signatures, see \cite[Example~9.4.11]{Rajaratnam2014}. Proceeding as in the above example and using additional results from \cite[Chapter~9]{Rajaratnam2014}, one can classify all (isometrically inequivalent) separable coordinates associated with Benenti tensors in~$\mathbb{E}^2$, including polar and Cartesian coordinates. The results of this classif\/ication are given in Table~\ref{tab:E2}. Benenti tensors in~$\mathbb{E}^2_1$, however, are richer, and so we introduce some theorems before classifying them in Section~\ref{sec:sepE21}.

\begin{table}[h]	\centering
	\caption{Separable coordinate systems in $\mathbb{E}^2$.}	\label{tab:E2}
	\begin{tabular}{| l | l | l |}
		\hline
		1) Cartesian coordinates & $L = d \odot d $ & $x d + y e$ \\
		2) polar coordinates & $L = r \odot r $ & $\rho \cos \theta d + \rho \sin \theta e$ \\
		3) elliptic coordinates & $L = d \odot d + a^{-2} r \odot r $ & $a\cos \phi \cosh \eta d + a\sin \phi \sinh \eta e$ \\
		4) parabolic coordinates & $L = 2 r \odot d $ & $\frac{1}{2}(\mu^{2} - \nu^{2}) d + \mu\nu e$ \\
		\hline
	\end{tabular} \\[1.5pt]
	The vectors $d$, $e$ form an orthonormal basis for $\mathbb{E}^2$ and $a > 0$.
\end{table}

\begin{figure}[h]
		\centering
	\begin{tikzpicture}
	\matrix (m) [matrix of nodes,column sep = 3mm,ampersand replacement=\&]
	{
		\node [geo] {$E_{1}$};
		\&
		\node [geo,color=white,text=black] {$\cdots$};
		\&
		\node [geo] {$E_{n}$}; \\
	};
	\end{tikzpicture}
\caption{Concircular tensor with simple eigenspaces $E_{1},\dots,E_{n}$.} \label{fig:Ldia}
\end{figure}
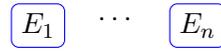

The following diagram of a Benenti tensor (see Fig.~\ref{fig:Ldia}) will be used later on. It represents the structure of the separable web associated with the Benenti tensor, which is the simplest possible. In the following section we will show how to use these webs to construct a richer class of separable webs called \emph{Kalnins--Eisenhart--Miller $($KEM$)$ webs}.

In general, an orthogonal concircular tensor may have multidimensional eigenspaces, and hence doesn't correspond to a separable web. But in two dimensions, all non-trivial\footnote{By a non-trivial concircular tensor, we mean one which is not a multiple of the metric when $n > 1$.} orthogonal CTs are Benenti tensors, which correspond to separable webs. Throughout the remainder of this section, we will classify all isometrically inequivalent separable webs in the two dimensional Minkowski space, $\mathbb{E}^2_1$, by studying their associated Benenti tensors.

We must f\/irst review the metric-Jordan canonical form of a self-adjoint operator on a pseudo-Euclidean space, $\mathbb{E}^{n}_{\nu}$ (i.e., a linear map $T$ on $\mathbb{E}^{n}_{\nu}$ such that $\langle T x, y\rangle = \langle x, T y\rangle $ for all $x,y \in \mathbb{E}^{n}_{\nu}$, where $\langle \cdot, \cdot\rangle$ is the scalar product). The details of the theory behind this canonical form are given in \cite[Appendix~C]{Rajaratnam2014}; these are solutions to Exercises~18 and~19 in \cite[pp.~260--261]{barrett1983semi}.

A \emph{Jordan block} of dimension $k$ with eigenvalue $\lambda \in \mathbb{C}$ is a $k \times k$ matrix denoted by $J_k(\lambda)$, and def\/ined as
\begin{gather*}
J_k(\lambda) :=
\begin{pmatrix}
\lambda & 1 & & & \\
& \lambda & \ddots & & 0 \\
& & \ddots & 1 & \\
& & & \lambda & 1 \\
& 0 & & & \lambda
\end{pmatrix}.
\end{gather*}

The \emph{skew-diagonal matrix} of dimension $k$ is denoted by $S_k$, and def\/ined as
\begin{gather*}
S_k := \begin{pmatrix}
0 & & 1 \\
& \ddots & \\
1 & & 0
\end{pmatrix}.
\end{gather*}

An ordered sequence of vectors $\beta = \{v_1,\dots,v_k\}$ where the matrix representation of $g$ with respect to (w.r.t.) $\beta$ has the form $g|_\beta = \varepsilon S_k$, is called a \emph{skew-normal sequence} of (length~$k$) and (sign $\varepsilon = \pm 1$). The subspace spanned by a skew-normal sequence is necessarily non-degenerate and of dimension $k$ (see \cite[Lemma~8.1.1]{Rajaratnam2014}).

In order to express the metric-Jordan canonical form of a self-adjoint operator on a pseudo-Euclidean space \cite[Appendix~C]{Rajaratnam2014}, we use the signed integer $\varepsilon k \in \mathbb{Z}$ where $k \in \mathbb{N}$ and $\varepsilon = \pm 1$. Then the notation $J_{\varepsilon k}(\lambda)$ is short hand for the pair:
\begin{gather*}
A = J_k(\lambda), \qquad g = \varepsilon S_k.
\end{gather*}

Furthermore, given matrices $A_1$ and $A_2$, we denote the following block diagonal matrix by $A_1 \oplus A_2$
\begin{gather*}
A_1 \oplus A_2 := \begin{pmatrix}
A_1 & 0 \\
0 & A_2
\end{pmatrix}.
\end{gather*}

The (real) metric-Jordan canonical form of a self-adjoint operator is discussed in detail in \cite[Appendix~C]{Rajaratnam2014}. In this article (for convenience) we will be working with the complex version (it can be deduced from \cite[Theorem~C.3.7]{Rajaratnam2014}), which is given as follows:

\begin{Theorem}[complex metric-Jordan canonical form \cite{barrett1983semi}] \label{thm:comMetJFor}
	A real operator $T$ on a pseudo-Euclidean space $\mathbb{E}^{n}_{\nu}$ is self-adjoint iff there exists a $($possibly complex$)$ basis $\beta$ such that
	\begin{gather*}
	T|_{\beta} = J_{\varepsilon_1 k_1}(\lambda_1) \oplus \cdots \oplus J_{\varepsilon_l k_l}(\lambda_l).
	\end{gather*}
Furthermore there exists a canonical basis such that the unordered list $\{J_{\varepsilon_1 k_1}(\lambda_1), \dots, J_{\varepsilon_l k_l}(\lambda_l) \}$ is uniquely determined by $T$ and an invariant of $T$ under the action of the orthogonal group~$O(\mathbb{E}^{n}_{\nu})$.
\end{Theorem}
\begin{Remark}
	Since $T$ is real, each Jordan block $J_{\varepsilon k}(\lambda)$ with $\lambda \in \mathbb{C} \setminus \mathbb{R}$ comes with a complex conjugate pair $J_{\varepsilon k}(\overline{\lambda})$. For complex eigenvalues, we can additionally assume that $\varepsilon = 1$.
\end{Remark}

A key fact used to derive the above canonical form and one to keep in mind is that for any self-adjoint operator $T$, any non-degenerate $T$-invariant subspace has a $T$-invariant orthogonal complement.

We are interested in classifying separable webs on a manifold $M$ modulo isometric equivalence. Since we are studying separable webs generated by Benenti tensors, we introduce the following notion of equivalence between CTs. Let $L$ be a CT in~$M$. We say a CT $\tilde{L}$ in $M$ is \emph{geometrically equivalent} to $L$ if there exists an isometry, $T$, of $M$, and constants $a \in \mathbb{R} \setminus {0}$ and $b \in \mathbb{R}$ such that
\begin{gather*}
\tilde{L} = a T_{*} L + b G.
\end{gather*}

It can be shown that if $L$ is a Benenti tensor which is not covariantly constant\footnote{In the covariantly constant case, the above statement doesn't hold, but this is not important for our purposes.}, then a~CT~$\tilde{L}$ is geometrically equivalent to~$L$ if\/f~$\tilde{L}$ is a Benenti tensor inducing a separable web isometrically equivalent to the one induced by~$L$ \cite[Proposition~6.2.5]{Rajaratnam2014}. Thus to classify isometrically inequivalent separable webs induced by Benenti tensors, we must classify geometrically inequivalent Benenti tensors.

We now review some general results from the classif\/ication of concircular tensors in $\mathbb{E}^{n}_{\nu}$ modulo geometric equivalence from \cite[Chapter~9]{Rajaratnam2014}. Let $L = A + 2 w \odot r + m r \odot r$ be the general concircular tensor in $\mathbb{E}^{n}_{\nu}$ def\/ined originally in equation~\eqref{eq:CTGenEunn}. For $k \geq 0$, def\/ine constants $\omega_k$ as follows:
\begin{gather*}
\omega_k =
\begin{cases}
m & \text{if} \ k = 0, \\
\big\langle w,A^{k-1} w\big\rangle & \text{else},
\end{cases}
\end{gather*}
where $\langle \cdot,\cdot \rangle$ denotes the pseudo-Euclidean scalar product. The above constants aren't necessarily invariant under isometries. But invariants can be def\/ined from them as follows.

\begin{Definition} \label{def:CtEunnInd}
Suppose $L$ is a CT in $\mathbb{E}^{n}_{\nu}$ as in equation~\eqref{eq:CTGenEunn}. Then we def\/ine the \emph{index} of $L$ to be the f\/irst integer $k \geq 0$ for which $\omega_k \neq 0$; $L$ is said to be \emph{non-degenerate} if such an integer exists. Furthermore if $L$ is non-degenerate, it has an associated sign (characteristic):
	\begin{gather*}
	\varepsilon =
	\begin{cases}
	1 & \text{if $k$ is even}, \\
	\operatorname{sgn} \omega_k & \text{if $k$ is odd}.
	\end{cases}
	\end{gather*}
\end{Definition}

The following theorem which is proven in \cite[Section~9.2]{Rajaratnam2014} summarizes our results on the canonical forms of concircular tensors; it classif\/ies C-tensors into f\/ive disjoint classes. In this theorem and its applications, the concircular tensor is considered to be a linear transformation of $\mathbb{E}^{n}_{\nu}$ into itself.
\begin{Theorem}[canonical forms for CTs in $\mathbb{E}^{n}_{\nu}$] \label{thm:conTenCanForm}
	Let $\tilde{L} = \tilde{A} + m r \otimes r^{\flat} + w \otimes r^{\flat} + r \otimes w^{\flat}$ be a~CT in $\mathbb{E}^{n}_{\nu}$. Let $k$ be the index and $\varepsilon$ be the sign of $\tilde{L}$ if $\tilde{L}$ is non-degenerate. These quantities are geometric invariants of $\tilde{L}$. Furthermore, after a possible change of origin and after changing to a geometrically equivalent CT, $L = a \tilde{L}$ for some $a \in \mathbb{R} \setminus \{0\}$, $\tilde{L}$ admits precisely one of the following canonical forms.
\begin{enumerate}\itemsep=0pt
		\item[]{\bf Central:} If $k=0$
		\begin{gather*}
		L = A + r \otimes r^{\flat}.
		\end{gather*}
		\item[]{\bf Non-null Axial:} If $k=1$, i.e., $m = 0$, and $\langle w,w\rangle \neq 0$:
			There exists a vector $e_1 \in \operatorname{span} \{w\}$ such that $L$ has the following form:
		\begin{gather*}
		L = A + e_{1} \otimes r^{\flat} + r \otimes e_{1}^{\flat},\qquad A e_{1} = 0, \qquad \langle e_{1},e_{1}\rangle = \varepsilon.
		\end{gather*}
		\item[] {\bf Null Axial:} If $k \geq 2$, hence $m = 0$ and $\langle w,w\rangle = 0$:
				There exists a skew-normal sequence $\beta = \{e_{1},\dots,e_{k}\}$ with $\langle e_{1},e_{k}\rangle = \varepsilon$ where $e_1 \in \operatorname{span} \{w\}$ which is $A$-invariant such that~$L$ has the following form:
		\begin{gather*}
		L = A + e_1 \otimes r^{\flat} + r \otimes e_1^{\flat} ,\qquad
		A|_{\beta} = J_k(0)^T =
		\begin{pmatrix}
		0 & & & & \\
		1 & 0 & & & \\
		& 1 & \ddots & & \\
		& & \ddots & 0 & \\
		& & & 1 & 0
		\end{pmatrix}.
		\end{gather*}
		\item[] {\bf Cartesian:} If $k$ doesn't exist, $m = 0$ and $w = 0$
		\begin{gather*}
		L = \tilde{A}.
		\end{gather*}
		\item[] {\bf Degenerate null Axial:} If $k$ doesn't exist and $w \neq 0$.
	\end{enumerate}
\end{Theorem}
\begin{Remark}
	The degenerate null axial concircular tensors will be of no concern to us. In Euclidean space they don't occur, and it can be shown that they are never orthogonal concircular tensors in Minkowski space (see \cite[Section~9.2.3]{Rajaratnam2014}).
\end{Remark}

One can easily deduce that in Euclidean or Minkowski space, any covariantly non-constant OCT is non-degenerate. Hence non-degenerate CTs are the main interest of this article. We now proceed to enumerate the isometrically inequivalent separable coordinates in $\mathbb{E}^2_1$.

\subsubsection[Separable coordinates in $\mathbb{E}^2_1$]{Separable coordinates in $\boldsymbol{\mathbb{E}^2_1}$} \label{sec:sepE21}

The simplest separable coordinate system in $\mathbb{E}^2_1$ is the Cartesian coordinate system (Case~1) which is generated by the Cartesian CT, $L = A$, where $A = \operatorname{diag}(\lambda_1, \lambda_2)$ with $\lambda_1 \neq \lambda_2$. In this case the eigenvalues of $A$ are geometrically insignif\/icant, and the separable coordinates are uniquely determined by the orthogonal eigenspaces of~$A$.

Throughout this classif\/ication we let $(t,x)$ denote the standard Cartesian coordinates on $\mathbb{E}^2_1$ with metric $g = \operatorname{diag}(-1,1)$, and the associated lightlike coordinates $(\zeta, \eta)$ are
\begin{gather*}
\zeta := \frac{1}{\sqrt{2}}(t - x), \qquad \eta := \frac{1}{\sqrt{2}}(t + x).
\end{gather*}

{\bf Central CTs.}
From the above theorem, the central CTs are the CTs with
\begin{gather*}
L = A + r \odot r.
\end{gather*}

Now we enumerate the (isometrically) inequivalent separable coordinates arising from central CTs by enumerating the inequivalent canonical forms for $A$.

{\bf Case 2, $\boldsymbol{A = 0}$.}
The eigenfunctions of $L$ are $0$ and $x^2 - t^2$ (with corresponding eigenspaces~$r^{\bot}$ and $\operatorname{span} \{r\}$), and~$L$ is (reducible) a Benenti tensor whenever~$|t| \neq|x|$, i.e., a dense subset of~$\mathbb{E}^2_1$, divided into four disjoint regions: the two disjoint timelike regions ($t > |x|$ and $t < -|x|$), and the two disjoint spacelike regions ($x > |t|$ and $x < -|t|$).

In the timelike regions, $L$ is not an ICT, hence a warped product\footnote{These are not needed now and will be def\/ined in the following section.} must be used to calculate the transformation formula.
However, one recognizes that these are the standard Rindler coordinates, whose transformation to Cartesian coordinates is given by $t = \pm u \cosh{v}$ and $x = u \sinh{v}$, where the $\pm$ applies to the appropriate region. The metric in these coordinates is
\begin{gather*}
ds^2 = -du^2 + u^2 dv^2.
\end{gather*}

Similarly in the spacelike regions, one recognizes these as Rindler coordinates, whose transformation to Cartesian coordinates is given by $t = u \sinh{v}$ and $x = \pm u \cosh{v}$, where the $\pm$ applies to the appropriate region. Here, the metric is
\begin{gather*}
ds^2 = du^2 - u^2 dv^2.
\end{gather*}

{\bf Case 3, $\boldsymbol{A = J_{-1}(\lambda_1) \oplus J_{1}(\lambda_2)}$ and $\boldsymbol{\lambda_1 < \lambda_2}$.} The characteristic polynomial $p(z)$ of $L$ can easily be calculated in Cartesian coordinates, yielding
\begin{gather} \label{eq:Case1CharPol}
p(z) = z^2 - \big(\lambda_1 + \lambda_2 + x^2 - t^2\big)z + \lambda_1\lambda_2 + \lambda_1x^2 -\lambda_2t^2.
\end{gather}
Since $L$ has no constant eigenfunctions, and it is therefore an ICT near any point where the eigenfunctions are simple as in Example~\ref{ex:ellipCoord}, one can show that $L$ has simple eigenfunctions in a dense subset of $\mathbb{E}^2_1$, namely where $t \neq 0$ and $x \neq 0$, and that the transformation from $(t,x)$ is given by
\begin{gather*} t^2 = \frac{(u - \lambda_1)(\lambda_1 - v)}{\lambda_2 - \lambda_1}, \qquad x^2 = \frac{(u - \lambda_2)(\lambda_2 - v)}{\lambda_2 - \lambda_1}.
\end{gather*}
We may, without loss of generality, let $u$ be the larger of the two eigenfunctions. Thus we have that $u$ and $v$ satisfy $v < \lambda_1 < \lambda_2 < u$. We may simplify these formulae by choosing instead to work with the geometrically equivalent concircular tensor, $\tilde{L} = L - \lambda_1 G$, and def\/ining new coordinates $(\bar{u}, \bar{v})$ by
\begin{gather*}
u = a^2 \cosh^2{\bar{u}}, \qquad v = -a^2 \sinh^2{\bar{v}},
\end{gather*}
where $a := \sqrt{\lambda_2 - \lambda_1} > 0$, and $0 \leq \bar{u}, \bar{v} < \infty$. The coordinate transformation from $(t,x)$ to the new coordinates $(\bar{u},\bar{v})$ is now given by the above equations upon the appropriate substitutions. The metric takes the Liouville form
\begin{gather*}
ds^2 = a^2\big(\cosh^2{\bar{u}}+\sinh^2{\bar{v}}\big)\big(d\bar{u}^2 - d\bar{v}^2\big).
\end{gather*}

{\bf Case 4, $\boldsymbol{A = J_{-1}(\lambda_1) \oplus J_{1}(\lambda_2)}$ and $\boldsymbol{\lambda_1 > \lambda_2}$.}
The characteristic polynomial is given again by equation~\eqref{eq:Case1CharPol}. One can calculate the discriminant of this polynomial to be $\Delta = 4\left( 2 \zeta^2 - a^2 \right) \left( 2 \eta^2 - a^2 \right)$, where $a := \sqrt{\lambda_1 - \lambda_2} > 0$. We see that $L$ has real simple eigenfunctions in f\/ive disjoint regions of~$\mathbb{E}^2_1$, which we label as follows
\begin{alignat*}{3}
& \textrm{N}\colon \quad 	&&\big\{(t,x) \in \mathbb{E}^2_1 \, | \, t - x > a, \ t + x > a \big\}, & \\
& \textrm{S}\colon \quad 	&&\big\{(t,x) \in \mathbb{E}^2_1 \, | \, t - x < -a, \ t + x < -a \big\}, &\\
&\textrm{E}\colon \quad 	&&\big\{(t,x) \in \mathbb{E}^2_1 \, | \, t - x < -a, \ t + x > a \big\}, &\\
&\textrm{W}\colon \quad 	&&\big\{(t,x) \in \mathbb{E}^2_1 \, | \, t - x > a, \ t + x < -a \big\}, & \\
&\textrm{C}\colon \quad 	&&\big\{(t,x) \in \mathbb{E}^2_1 \, | \, | t - x| < a, \ |t + x| < a\big\}. &
\end{alignat*}

The transformation from canonical coordinates $(u,v)$ to Cartesian coordinates are calculated as in the previous case, and are given as follows
\begin{gather*}
t^2 = \frac{(\lambda_1 - u)(\lambda_1 - v)}{\lambda_1 - \lambda_2}, \qquad x^2 = \frac{(\lambda_2 - u)(\lambda_2 - v)}{\lambda_1 - \lambda_2}.
\end{gather*}

WLOG, we take $u$ to be the larger of the two eigenfunctions, and $\lambda_2 = 0$ after passing to a geometrically equivalent CT. Then the above observations imply that $u$ and $v$ satisfy $v < u < 0 < a^2$ in regions N and S, $0 < a^2 < v < u$ in regions~E and~W, and $0 < v < u < a^2$ in region C. Noting the domain of the coordinates $(u,v)$, we def\/ine new coordinates $(\bar{u},\bar{v})$ in each respective region by
\begin{alignat*}{4}
& \textrm{N and S}\colon \quad	&& u = -a^2 \sinh^2{\bar{u}},\qquad && v = -a^2 \sinh^2{\bar{v}}, & \\
& \textrm{E and W}\colon \quad	&& u = a^2 \cosh^2{\bar{u}},\qquad && v = a^2 \cosh^2{\bar{v}}, & \\
& \textrm{C}\colon \quad		&& u = a^2 \sin^2{\bar{u}},\qquad && v = a^2 \sin^2{\bar{v}}, &
\end{alignat*}
where $0 < \bar{v} < \bar{u} < \infty$ in regions~N and~S, $0 < \bar{u} < \bar{v} < \infty$ in regions~E and~W, and $0 < \bar{v} < \bar{u} < \frac{\pi}{2}$ in region C. The transformation from $(\bar{u},\bar{v})$ to Cartesian coordinates is given by the above equations after the appropriate substitution. The metric takes the following forms in the various regions:
\begin{alignat*}{3}
&\textrm{N and S}\colon \quad	&& ds^2 = a^2\big(\cosh^2{\bar{u}} - \cosh^2{\bar{v}}\big)\big(d\bar{u}^2 - d\bar{v}^2\big),& \\
&\textrm{E and W}\colon \quad	&& ds^2 = a^2\big(\cosh^2{\bar{v}} - \cosh^2{\bar{u}}\big)\big(d\bar{u}^2 - d\bar{v}^2\big),& \\
&\textrm{C}\colon \quad		&& ds^2 = a^2\big(\cos^2{\bar{u}} - \cos{\bar{v}}\big)\big(d\bar{u}^2 - d\bar{v}^2\big).& \\
\end{alignat*}

{\bf Case 5, $\boldsymbol{A = J_{1}(\alpha +ib) \oplus J_{1}(\alpha -i b)}$.}
We f\/irst note that we can assume $b > 0$ (resp. $\alpha = 0$) after applying an isometry (resp.\ transforming to a geometrically equivalent CT) if necessary. Then the characteristic polynomial and discriminant of $L$ are
\begin{gather*}
p(z) = z^2 - \big(x^2 - t^2\big)z + b^2 + 2btx, \qquad \Delta = \big((t-x)^2 - 2b\big)\big((t+x)^2 + 2b\big).
\end{gather*}
Therefore, $L$ has real simple eigenfunctions only in the regions with $|\zeta| > \sqrt{b}$, where $(\zeta, \eta)$ are the standard lightlike coordinates. The transformation from the lightlike coordinates to canonical separable coordinates $(u,v)$ is given by
\begin{gather*}u + v = -2\zeta \eta,\qquad (u - v)^2 = 4\big(\eta^2 + b\big)\big(\zeta^2 - b\big).\end{gather*}
We may def\/ine new coordinates $(\bar{u}, \bar{v})$ by $u = b \sinh{2\bar{u}}$ and $v = -b \sinh{2\bar{v}}$, where now $\bar{u} > |\bar{v}|$. Then the transformation formulae are equivalent to
\begin{gather*}\zeta = \pm \sqrt{b} \cosh{(\bar{u} + \bar{v})}, \qquad \eta = \mp \sqrt{b} \sinh{(\bar{u} - \bar{v})},\end{gather*}
 where the $\pm$ applies in the region where $\zeta \gtrless \pm \sqrt{b}$. The metric then takes the following form (where we have def\/ined $a := \sqrt{b}$):
\begin{gather*} ds^2 = a^2\big(\sinh{2\bar{u}} + \sinh{2\bar{v}}\big)\big(d\bar{u}^2 - d\bar{v}^2\big).\end{gather*}

{\bf Case 6, $\boldsymbol{A = J_{-2}(\lambda)}$.} We f\/irst note that we can assume $\lambda = 0$ after passing to a geometrically equivalent CT. Then the characteristic polynomial and discriminant of $L$ in the associated orthogonal coordinates $(t,x)$ are
\begin{gather*}
p(z) = z^2 - \big(x^2 - t^2\big)z - \tfrac{1}{2}(x-t)^2, \qquad \Delta = \big(x^2 - t^2\big)^2 + 2(x-t)^2.
\end{gather*}
From the discriminant, $L$ has real simple eigenfunctions, and hence induces an ICT, everywhere except on the line $x = t$. The transformation from the lightlike coordinates to canonical separable coordinates $(u,v)$ is given by
\begin{gather*} u+ v = -2\zeta \eta, \qquad (u - v)^2 = 4\zeta^2 \big(\eta^2 + 1\big)\end{gather*}
with $- \infty < v < 0 < u < \infty$. Noting the constraints, we def\/ine $(\bar{u},\bar{v})$ by $u = e^{2\bar{u}}$, and $v = -e^{2\bar{v}}$, with $- \infty < \bar{u}, \bar{v} < \infty$. Then the transformation formulae are equivalent to
\begin{gather*} \zeta = \pm e^{\bar{u} + \bar{v}}, \qquad \eta = \sinh{(\bar{u} - \bar{v})}, \end{gather*}
where the $\pm$ applies in regions $\zeta \gtrless 0$ respectively. The metric then takes the form
\begin{gather*} ds^2 = \big(e^{2\bar{u}} + e^{2\bar{v}}\big)\big(d\bar{u}^2 - d\bar{v}^2\big).\end{gather*}

{\bf Case 7, $\boldsymbol{A = J_{2}(\lambda)}$.}
We f\/irst note that we can assume $\lambda = 0$ after passing to a geometrically equivalent CT. Then the characteristic polynomial and discriminant of $L$ in the associated orthogonal coordinates $(t,x)$ are
\begin{gather*}
p(z) = z^2 - \big(x^2 - t^2\big)z + \tfrac{1}{2}(x+t)^2,\qquad \Delta = 4\zeta^2\big(\eta^2 - 1\big).
\end{gather*}
Thus $L$ has distinct real eigenfunctions in the regions $|\eta| > 1$, except on the line $\zeta = 0$. Furthermore, the eigenfunctions of $L$ are
\begin{gather*}
u = \eta \zeta + \sqrt{\zeta^2\big(\eta^2 - 1\big)},\qquad v = \eta \zeta - \sqrt{\zeta^2\big(\eta^2 - 1\big)}.
\end{gather*}
The regions in which $L$ is an ICT is divided into four disjoint subsets which we label:
\begin{alignat*}{3}
& \textrm{N}\colon \quad &&\big\{(\eta,\zeta) \in \mathbb{E}^2_1 \, | \, \eta > 1, \ \zeta < 0 \big\}, & \\
& \textrm{S}\colon \quad &&\big\{(\eta,\zeta) \in \mathbb{E}^2_1 \, | \, \eta < -1, \ \zeta > 0 \big\}, & \\
& \textrm{E}\colon \quad &&\big\{(\eta,\zeta) \in \mathbb{E}^2_1 \, | \, \eta > 1, \ \zeta > 0 \big\}, & \\
& \textrm{W}\colon \quad &&\big\{(\eta,\zeta) \in \mathbb{E}^2_1 \, | \, \eta < -1, \ \zeta < 0 \big\}, &
\end{alignat*}
where N and S are timelike regions, and E and W are spacelike. The above observations imply that $u$ and $v$ satisfy $v < u < 0$ in regions N and S, and $0 < v < u$ in regions E and W. We can therefore def\/ine, in each region, new coordinates $\bar{u}$ and $\bar{v}$ by $u = \pm e^{2\bar{u}}$, and $v = \pm e^{2\bar{v}}$, with the~$\pm$ corresponding to the appropriate sign of $u$ and $v$, satisfying $\bar{v} < \bar{u}$ in regions E and W and $\bar{u} < \bar{v}$ in regions N and S. Thus we have that the transformation formulae are equivalent to
\begin{gather*}\eta = \pm \cosh{(\bar{u} - \bar{v})},\qquad \zeta = \pm e^{\bar{u} + \bar{v}}, \end{gather*}
where the $\pm$ in each equation applies where the respective coordinate has the appropriate sign. The metric in takes the following forms in the various regions:
\begin{alignat*}{3}
& \textrm{N and S}\colon \quad && ds^2 = \big(e^{2\bar{v}} - e^{2\bar{u}}\big)\big(d\bar{u}^2 - d\bar{v}^2\big),& \\
& \textrm{E and W}\colon \quad && ds^2 = \big(e^{2\bar{u}} - e^{2\bar{v}}\big)\big(d\bar{u}^2 - d\bar{v}^2\big).&
\end{alignat*}

{\bf Axial CTs.} From the above theorem, the axial CTs are CTs of the form
\begin{gather*}
L = A + w \odot r + r \odot w.
\end{gather*}

Now we enumerate the isometrically inequivalent separable coordinates associated with axial CTs by enumerating the (geometrically) inequivalent canonical forms for the pair $(A,w)$ given by above theorem.

{\bf Non-null Axial CTs: Case~8, $\boldsymbol{A = 0}$ and $\boldsymbol{\langle w,w\rangle = -1}$.}
We may choose our Cartesian coordinates so that $w = \partial_t$. The characteristic polynomial of $L$ is $p(z) = (z + t)^2 - t^2 + x^2$. Thus we see that $L$ has real simple eigenfunctions where $t^2 > x^2$, i.e., in the two timelike regions ($t > |x|$ and $t < -|x|$). The transformation equations are given by
\begin{gather*} x^2 = uv, \qquad t = -\tfrac{1}{2}(u+v)\end{gather*}
with $- \infty < v < u < 0$ for $t > 0$, and $0 < v < u < \infty$ for $t < 0$. We may introduce new coordinates $(\bar{u},\bar{v})$ def\/ined by $(u,v) = (-\bar{v}^2,-\bar{u}^2)$ for $t > 0$, and by $(u,v) = (\bar{u}^2,\bar{v}^2)$ for $t < 0$. We therefore have $0 < \bar{v} < \bar{u} < \infty$, and the transformation formulae are easily obtained from the above. The metric is in these coordinates takes the form
\begin{gather*}ds^2 = \big(\bar{u}^2 - \bar{v}^2\big)\big(d\bar{v}^2 - d\bar{u}^2\big).\end{gather*}

{\bf Non-null Axial CTs: Case 9, $\boldsymbol{A = 0}$, and $\boldsymbol{\langle w,w\rangle = 1}$.}
This case is analogous to the one above. We choose our Cartesian coordinates so $w = \partial_x$. The characteristic polynomial of~$L$ is $p(z) = (z - x)^2 - x^2 + t^2$. So $L$ has real simple eigenfunctions where $x^2 > t^2$, i.e., in the two spacelike regions ($x > |t|$ and $x < -|t|$). The transformation from coordinates $(u,v)$ to Cartesian coordinates are
\begin{gather*} t^2 = uv,\qquad x = \tfrac{1}{2}(u + v)\end{gather*}
\noindent with $0 < v < u < \infty$ where $x > 0$, and $- \infty < v < u < 0$ where $x < 0$. We may introduce new coordinates $(\bar{u},\bar{v})$ def\/ined by $(u,v) = (\bar{u}^2,\bar{v}^2)$ for $x > 0$ and by $(u,v) = (-\bar{v}^2,-\bar{u}^2)$ for $x < 0$. Hence we have $0 < \bar{v} < \bar{u} < \infty$. The transformation formulae are easily obtained from the above equations, and the metric in these coordinates takes the form
\begin{gather*} ds^2 = \big(\bar{u}^2 - \bar{v}^2\big)\big(d\bar{u}^2 - d\bar{v}^2\big).\end{gather*}

{\bf Null Axial CTs: Case 10, $\boldsymbol{A = J_{-2}(0)}$ and $\boldsymbol{\langle w,w\rangle = 0}$.}
We may choose our Cartesian coordinates so that $w = \partial_\eta$. Notice that the null axial CT with $A = J_{2}(0)$ is geometrically equivalent to this one, after multiplying $L$ by $-1$. In the associated Cartesian coordinates $(t,x)$, the characteristic polynomial of $L$ is
\begin{gather*} p(z) = z^2 + \sqrt{2}(t + x)z -\sqrt{2}(t-x) + 2tx + \tfrac{1}{2}(t - x)^2\end{gather*}
with discriminant $\Delta = 4 \sqrt{2}(t-x) = 8\zeta$. Therefore, $L$ has distinct real eigenfunctions in the region $\zeta > 0$. The transformation from canonical coordinates $(u,v)$ to $(t,x)$ are
\begin{gather*} \eta = -\tfrac{1}{2}(u+v), \qquad \zeta = \tfrac{1}{8}(u-v)^2\end{gather*}
 with $u > v$. In these coordinates the metric takes the Liouville form
\begin{gather*}ds^2 = \tfrac{1}{4}(u - v)\big({-}du^2 + dv^2\big).\end{gather*}

This classif\/ication of separable coordinates in $\mathbb{E}^2_1$ is exhaustive due to the KEM separation theorem \cite[Theorem~1.4]{Rajaratnam2014d} (see also Theorem~\ref{Intthm:SCCconChKT}), which says: in $\mathbb{E}^2_1$ any separable coordinate system admits a non-trivial Benenti tensor which is diagonalized in the coordinates. The above results agree with those obtained earlier by direct integration of the EIC \cite{Kalnins1975, McLenaghan2002a}. See~\cite{Chanu2006} for a~dif\/ferent algebraic classif\/ication.

The classif\/ication of separable coordinates in $\mathbb{E}^2_1$ gives a clear picture of what is involved in the more general classif\/ication in $\mathbb{E}^n_1$. In $\mathbb{E}^n_1$ more possibilities arise. For example, if $\lambda_1 < \dots <\lambda_n$, then one can construct $n$ geometrically inequivalent central CTs, $L = A + r \otimes r^\flat$, from these parameters, where $A$ is given as follows
\begin{gather*}
A = J_{-1}(\lambda_1) \oplus J_{1}(\lambda_2) \oplus \cdots \oplus J_{1}(\lambda_n), \\
 \cdots\cdots\cdots\cdots\cdots\cdots\cdots\cdots\cdots \cdots\cdots\cdots\\
A = J_{1}(\lambda_1) \oplus \cdots \oplus J_{1}(\lambda_{n-1}) \oplus J_{-1}(\lambda_n).
\end{gather*}

\looseness=-1 We note that general formulas for the characteristic polynomial of OCTs and the metric of ICTs def\/ined in~$\mathbb{E}^n_1$ can be found in \cite[Section~9.4]{Rajaratnam2014}. Additionally, in higher dimensions, all non-trivial OCTs are not Benenti tensors. When an OCT has a multidimensional eigenspace, warped products can be used to build separable coordinate systems. We discuss this procedure in the following section, and then enumerate the isometrically inequivalent separable coordinates in~$\operatorname{dS}_2$.

\begin{figure}[t]\centering
		\begin{subfigure}{0.3\textwidth}
			\includegraphics[scale=0.3]{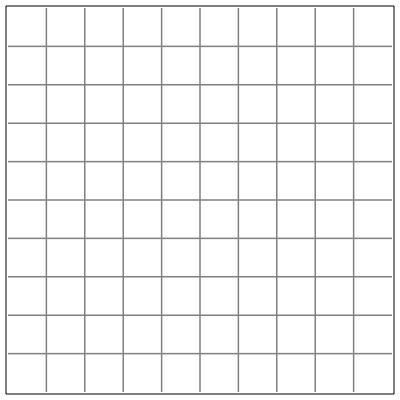}
			\caption{Cartesian}
			\label{fig:M2.1}
		\end{subfigure}
		\begin{subfigure}{0.3\textwidth}
			\includegraphics[scale=0.3]{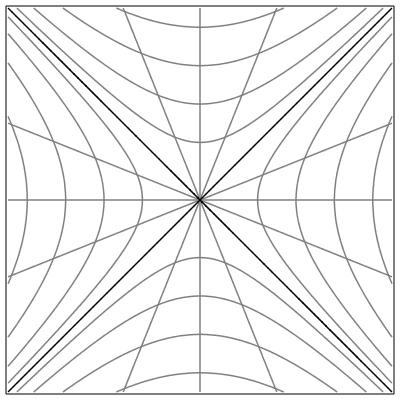}
			\caption{Rindler}
			\label{fig:M2.2}
		\end{subfigure}
		\begin{subfigure}{0.3\textwidth}
			\includegraphics[scale=0.3]{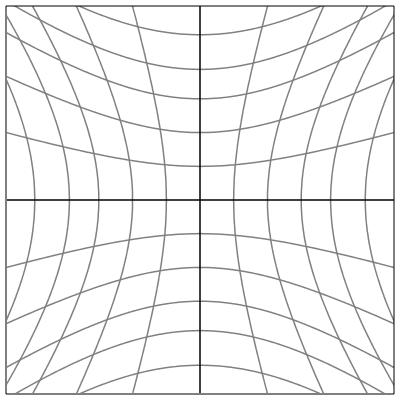}
			\caption{Real elliptic I}
			\label{fig:M2.3}
		\end{subfigure} \\
		\begin{subfigure}{0.3\textwidth}
			\includegraphics[scale=0.3]{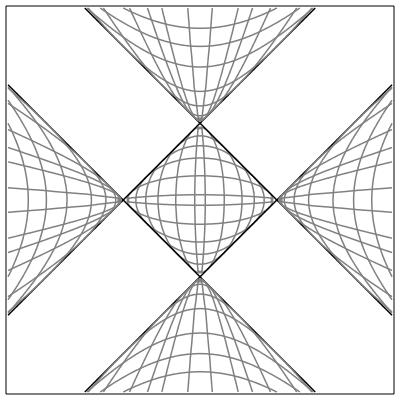}
			\caption{Real elliptic II}
			\label{fig:M2.4}
		\end{subfigure}
		\begin{subfigure}{0.3\textwidth}
			\includegraphics[scale=0.3]{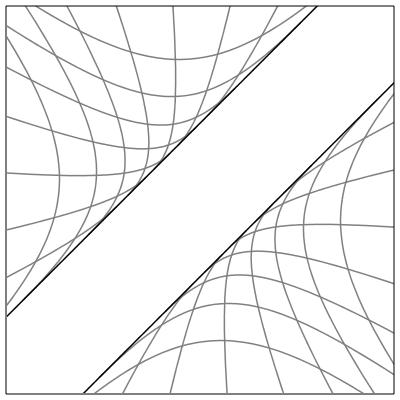}
			\caption{Complex elliptic}
			\label{fig:M2.5}
		\end{subfigure}
		\begin{subfigure}{0.3\textwidth}
			\includegraphics[scale=0.3]{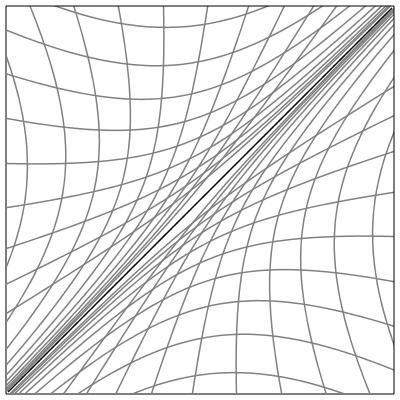}
			\caption{Null elliptic I}
			\label{fig:M2.6}
		\end{subfigure} \\
		\begin{subfigure}{0.3\textwidth}
			\includegraphics[scale=0.3]{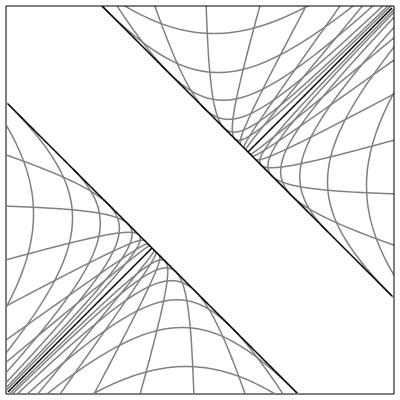}
			\caption{Null elliptic II}
			\label{fig:M2.7}
		\end{subfigure}
		\begin{subfigure}{0.3\textwidth}
			\includegraphics[scale=0.3]{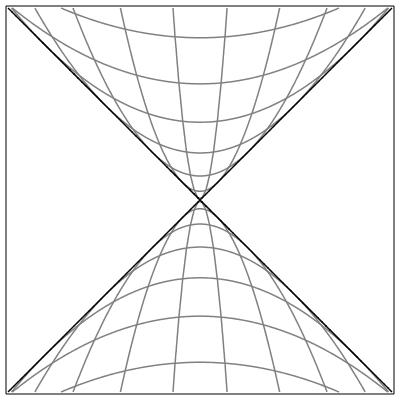}
			\caption{Timelike parabolic}
			\label{fig:M2.8}
		\end{subfigure}
		\begin{subfigure}{0.3\textwidth}
			\includegraphics[scale=0.3]{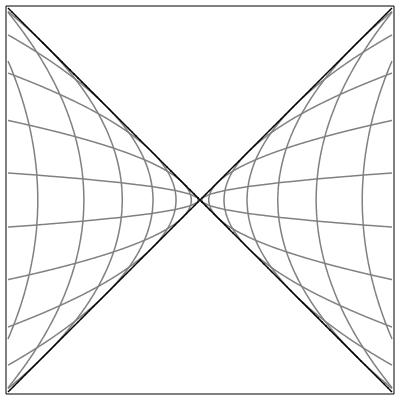}
			\caption{Spacelike parabolic}
			\label{fig:M2.9}
		\end{subfigure} \\
		\begin{subfigure}{0.3\textwidth}
			\includegraphics[scale=0.3]{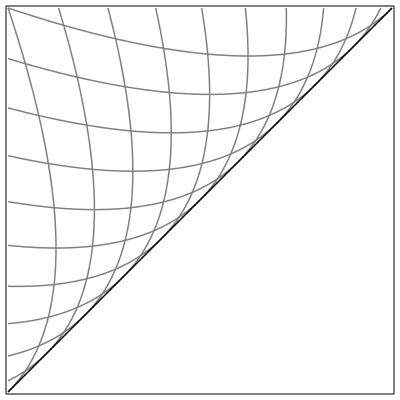}
			\caption{Null parabolic}
			\label{fig:M2.10}
		\end{subfigure}
\caption{Separable coordinate webs on $\mathbb{E}^2_1$.}		\label{fig:M2webs}
	\end{figure}

Having illustrated in detail the use of CTs in classifying the separable coordinate systems in~$\mathbb{E}^2_1$, we now tabulate the results obtained above as a reference. In the following table, we list the separable webs, their transformation equations, metrics and coordinate ranges, as well as (a~canonical choice for) the associated CT. Note that while we often give the transformation equations for a particular chart domain, the corresponding equation for all other chart domains can be obtained by isometry (for instance, some combination of $t \rightarrow \pm t$ and $x \rightarrow \pm x$). In the following, $e_0$ is a~unit timelike vector orthogonal to the spacelike unit vector~$e_1$, and $k$ is a~nonzero null vector. Following Table~\ref{tab:M2}, we present some graphics in Fig.~\ref{fig:M2webs}, obtained using Maple, which illustrate the separable webs for each case. These graphics may also be found in~\cite{Chanu2006}. In Fig.~\ref{fig:M2webs}, the empty white spaces containing no coordinate curves represent the open singular sets of the web, and the black lines represent the closed singular sets (in this case, singular lines) of the web.

	\begin{center}
		
		\begin{longtable}{|l|l|}
			\caption{Separable coordinate systems in $\mathbb{E}^2_1$.} \label{tab:M2} \\
			\hline
			\textbf{Cartesian CTs, $\boldsymbol{L = A}$} & \\
			\hline
			\textbf{Case 1:} 							& $ds^2 = -du^2 + dv^2$ \tsep{2pt}\\
			Cartesian coordinates						& $t = u$, $x = v$ \\
			$L = e_0 \odot e_0$						& $-\infty < u < \infty$, $-\infty < v < \infty$ \\
			\hline
			\textbf{Central CTs, $\boldsymbol{L = A + r \odot r}$} & \\
			\hline
			\textbf{Case 2:} 							&for $-t^2 + x^2 > 0$\tsep{2pt} \\
			Rindler coordinates							& $ds^2 = du^2 - u^2dv^2$ \\
			$L = r \odot r$						 	& $t = u \sinh{v}$, $x = u \cosh{v}$ \\
			& $0 < u < \infty$, $-\infty < v < \infty$ \bsep{3pt}\\
			&for $-t^2 + x^2 < 0$ \\
			& $ds^2 = -du^2 + u^2dv^2$ \\
			& $t = u \cosh{v}$, $x = u \sinh{v}$ \\
			& $0< u < \infty$, $-\infty < v < \infty$ \\
			\hline
			\textbf{Case 3:} & $ds^2 = a^2(\cosh^2{u} + \sinh^2{v})(du^2 - dv^2)$\tsep{2pt} \\
			Real elliptic coordinates of Type I 					& $t = a \cosh{u} \sinh{v}$, $x = a \cosh{v} \sinh{u}$ \\
			$L = a^2 e_1 \odot e_1 + r \odot r$, $a > 0$	& $0< u < \infty$, $0 < v < \infty$ \\
			\hline
			\textbf{Case 4:} 										& for $|t| - |x| > a$ \\
			Real elliptic coordinates of Type II					&$ds^2 = a^2(\cosh^2{u} - \cosh^2{v})(du^2 - dv^2)$ \\
			$L = a^2 e_0 \odot e_0 + r \odot r$, $a > 0$ & $t = a \cosh{u} \cosh{v}$, $x = a \sinh{v} \sinh{u}$ \\
			& $0 < v < u < \infty$ \bsep{3pt}\\
			& for $|t| - |x| < - a$ \\
			& $ds^2 = a^2(\cosh^2{v} - \cosh^2{u})(du^2 - dv^2)$ \\
			& $t = a \sinh{u} \sinh{v}$, $x = a \cosh{v} \cosh{u}$ \\
			& $0< u < v < \infty$ \bsep{3pt}\\
			&for $|t| + |x| < a$ \\
			& $ds^2 = a^2(\cos^2{u} - \cos^2{v})(du^2 - dv^2)$ \\
			& $t = a \cos{u} \cos{v}$, $x = a \sin{v} \sin{u}$ \\
			& $0< v < u < \frac{\pi}{2}$ \\
			\hline
			\textbf{Case 5:} 									& $ds^2 = a^2(\sinh{2u} + \sinh{2v})(du^2 - dv^2)$\tsep{2pt} \\
			Complex elliptic coordinates							& $t-x = \sqrt{2} a\cosh{(u + v)}$, $t+x = \sqrt{2} a\sinh{(u - v)}$ \\
			$L = a^2 e_0 \odot e_1 + r \odot r$, $a > 0$	& $0 < |v| < u < \infty$ \\
			\hline
			\textbf{Case 6:} 									& $ds^2 = (e^{2u} + e^{2v})(du^2 - dv^2)$\tsep{2pt} \\
			Null Elliptic Coordinates of Type I						& $t-x = \sqrt{2}e^{u+v}$, $t+x = \sqrt{2}\sinh{(u-v)}$ \\
			$L = -k \odot k + r \odot r$							& $-\infty < u < \infty$, $-\infty < v < \infty$ \\
			\hline
			\textbf{Case 7:} 							&for $-t^2 + x^2 > |t - x|$\tsep{2pt} \\
			Null elliptic coordinates of Type II				& $ds^2 = (e^{2u} - e^{2v})(du^2 - dv^2)$ \\
			$L = k \odot k + r \odot r$				 	& $t-x = -\sqrt{2}e^{u+v}$, $t+x = \sqrt{2}\cosh{(u-v)}$ \\
			& $-\infty < v < u < \infty$\bsep{3pt} \\
			&for $-t^2 + x^2 < -|t - x|$ \\
			& $ds^2 = (e^{2v} - e^{2u})(du^2 - dv^2)$ \\
			& $t-x = \sqrt{2}e^{u+v}$, $t+x = \sqrt{2}\cosh{(u-v)}$ \\
			& $-\infty < u < v < \infty$ \\
			\hline
			\textbf{Axial CTs, $\boldsymbol{L = A + 2w \odot r}$} & \\
			\hline
			\textbf{Case 8:} 									& $ds^2 = (u^2 - v^2)(-du^2+ dv^2)$\tsep{2pt} \\
			Timelike parabolic coordinates						& $t = \frac{1}{2}(u^2 + v^2)$, $x = uv$ \\
			$L = 2e_0 \odot r$									& $0 < v < u < \infty$ \\
			\hline
			\textbf{Case 9:} 									& $ds^2 = (u^2 - v^2)(du^2 - dv^2)$\tsep{2pt} \\
			Spacelike parabolic coordinates						& $t = uv$, $x = \frac{1}{2}(u^2 + v^2)$ \\
			$L = 2e_1 \odot r$									& $0 < v < u < \infty$ \\
			\hline
			\textbf{Case 10:} 									& $ds^2 = \frac{1}{4}(u - v)(-du^2 + dv^2)$ \tsep{2pt} \\
			Null parabolic coordinates							& $t+x = -\frac{\sqrt{2}}{2}(u+v)$, $t-x = \frac{\sqrt{2}}{8}(u - v)^2$ \\
			$L = 2k \odot r$									& $0 < v < u < \infty$ \\
			\hline
		\end{longtable}
	\end{center}

\subsection{Concircular tensors with multidimensional eigenspaces and KEM webs} \label{sec:CTsMultEig}

More generally, one can attempt to construct separable webs using any non-trivial orthogonal concircular tensor, as we will see in this section.
Suppose $L$ is a non-trivial\footnote{A CT is called non-trivial if its not a multiple of the metric.} orthogonal concircular tensor with a single multidimensional eigenspace $D$; denote by $D^{\perp}$ the distribution orthogonal to $D$. Then one can show that (see \cite[Theorem~6.1]{Rajaratnam2014a}):
\begin{itemize}\itemsep=0pt
	\item There is a local product manifold $B \times F$ of (pseudo-)Riemannian manifolds $(B, g_{B})$ and $(F, g_{F})$ such that:
	$\{p\} \times F$ is an integral manifold of $D$ for any $p \in B$ and $B \times \{q\}$ is an integral manifold of $D^{\perp}$ for any $q \in F$.
	\item $B \times F$ equipped with the metric $\pi_{B}^{*} g_{B} + \rho^{2} \pi_{F}^{*} g_{F}$ for a specif\/ic function $\rho\colon B \rightarrow \mathbb{R}^{+}$ is locally isometric to $(M,g)$.
\end{itemize}

Such a product manifold is called a \emph{warped product} and is denoted $B \times_{\rho} F$. The manifold $B$ is called the \emph{geodesic factor} and $F$ is called the \emph{spherical factor} of the warped product. We also say that the warped product $B \times_{\rho} F$ is \emph{adapted} to the splitting $(D^{\perp},D)$, which is often called a \emph{warped product net $($WP-net$)$}. When a distribution~$D$ admits an adapted warped product as above, it is called a \emph{Killing distribution}. See~\cite{Rajaratnam2014a} for more details on these matters.

We note here that warped products are rigid. For example, in Euclidean space, it can be shown (e.g., see~\cite{Nolker1996}) that if an open connected subset $U$ is isometric to a warped product with a single spherical factor, then the warped product must have one of the following forms:
\begin{enumerate}\itemsep=0pt
	\item[1)] $\mathbb{E}^{m} \times_{\rho} \mathbb{S}^{r}$,
	\item[2)] $\mathbb{E}^{m} \times_{1} \mathbb{E}^{r}$.
\end{enumerate}

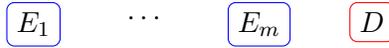
\begin{figure}[h]\centering
	\begin{tikzpicture}
	[level 1/.style={level distance=0mm},
	level 2/.style={level distance=1.5cm}]
	\coordinate
	child {node [geo] {$E_{1}$}
		edge from parent[draw=none] }
	child {node [geo,color=white,text=black] {$\cdots$}
		edge from parent[draw=none] }
	child {node [geo] {$E_{m}$}
		edge from parent[draw=none] }
	child {node [sph] {$D$}
		edge from parent[draw=none] };
	\end{tikzpicture}	\caption{Concircular tensor with eigenspaces $E_{1},\dots,E_{m},D$.} \label{fig:LDdia}
\end{figure}

Now, if we enumerate the one dimensional eigenspaces of $L$ by $E_1,\dots,E_m$ and denote the multidimensional eigenspace of $L$ by $D$ as above, then Fig.~\ref{fig:LDdia} gives a diagram for~$L$. In this f\/igure, the block containing the eigenspace~$D$ represents a ``degeneracy'' which needs to be removed to uniquely specify a separable web. We now describe how to do this.

A remarkable property of the warped product decomposition is the following. Let $\tilde{K}$ be a~ChKT on $F$, this can be canonically lifted to a~tensor, $\tilde{K} \in S^2(B \times_{\rho} F)$, which is in fact a~KT on~$B \times_{\rho} F$! Hence if $K'$ is the KBDT associated with~$L$, then locally we can assume that $K' + \tilde{K}$ is a ChKT on~$B \times_{\rho} F$. Indeed, one can show that~$L$ induces a Benenti tensor,~$L'$, on~$B$ by restriction. Let $(x^i)$ be any coordinates on~$B$ which diagonalize~$L'$. Note that we observed in the previous section that these coordinates are separable on $B$. Suppose $(y^j)$ are coordinates on $F$ which diagonalize $\tilde{K}$, hence are separable (see Theorem~\ref{Intthm:HJosepI}). Then since the product coordinates $(x^i , y^j)$ diagonalize $K' + \tilde{K}$ (see equation~\eqref{introeq:KTsumKEM}), Theorem~\ref{Intthm:HJosepI} implies that $K' + \tilde{K}$ is a ChKT\footnote{The eigenfunctions may not exactly be simple, but one can add a constant multiple of the metric on $F$ to $\tilde{K}$ so that they are locally simple.} and that these coordinates are separable. Note that in these coordinates $K' + \tilde{K}$ have the following form:
\begin{gather} \label{introeq:KTsumKEM}
K' + \tilde{K} = \sum\limits_{i} (\operatorname{tr}(L) - \lambda_i) \partial_{i} \otimes {\rm d} x^{i} + \sum\limits_{j} \big(\operatorname{tr}(L) - c + \tilde{\lambda}_j\big) \partial_{j} \otimes {\rm d} y^{j},
\end{gather}
 where $\lambda_i$ are the eigenfunctions of $L'$, $c$ is the constant eigenfunction of~$L$ associated with $D$ and $\tilde{\lambda}_j$ are the eigenfunctions of $\tilde{K}$. In conclusion, we have shown how to construct separable coordinates $(x^i , y^j)$ using the CT~$L$ and ChKT $\tilde{K}$. In fact, the entire Killing--St\"ackel space of the associated separable web can be calculated using the warped product \cite[Proposition~4.9]{Rajaratnam2014a}:

\begin{Proposition}[the Killing--St\"{a}ckel space of a reducible separable web] \label{prop:wpKss}
Suppose $J$ is a~ChKT in an arbitrary pseudo-Riemannian manifold, with associated KS-space $\mathcal{K}$ inducing a~reducible separable web, i.e., there exists a $J$-invariant Killing distribution $D$. Let $M = B \times_{\rho} F$ be a~local warped product adapted to the WP-net $(D^{\perp},D)$ with adapted contravariant metric $G = G_{B} + \rho^{-2} G_{F}$. Then there are KS-spaces $\mathcal{K}_{B}$ and $\mathcal{K}_{F}$ on $B$ and $F$ respectively such that $K \in \mathcal{K}$ iff there exists $K_{B} \in \mathcal{K}_{B}$, $K_{F} \in \mathcal{K}_{F}$ and $k \in \hat{\mathcal{F}}(B)$, where $\hat{\mathcal{F}}(B)$ denotes the pull back of~$\mathcal{F}(B)$ to~$\mathcal{F}(M)$ using the product decomposition, such that the following equations hold
	\begin{gather*}
	K = K_{B} + k G_{F} + K_{F}, \qquad 	{\rm d} k = K_{B} {\rm d} \rho^{-2}.
	\end{gather*}
\end{Proposition}

More generally, the above two equations characterize all Killing tensors $K$, for which $D$ is an invariant distribution, in terms of Killing tensors $K_{B}$ and $K_F$ on $B$ and $F$ respectively \cite[Proposition~4.3]{Rajaratnam2014a}.

On $B$, equations~\eqref{introeq:BTLeq} and \eqref{introeq:BTLseqb} gives a basis for the KS-space, $\mathcal{K}_{B}$, associated with $L'$. Using the above result, one can calculate the lifts of this basis to be \cite[Proposition~6.6.3]{Rajaratnam2014}:
\begin{gather*}
\bar{K}_{a} := K_{a} + \left(\sum_{i=0}^{a} (-c)^{i}\sigma_{a-i}\right)\rho^{-2} G_F, \qquad \sigma_{i} = \frac{1}{i}\operatorname{tr}(K_{i-1}L'),
\end{gather*}
where $0 \leq a \leq m -1$. One can now calculate the entire KS-space, $\mathcal{K}$, straightforwardly from the above proposition.

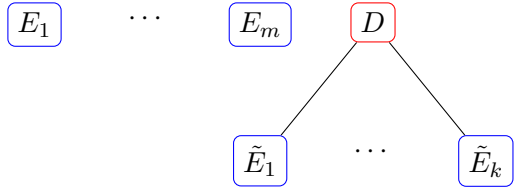
\begin{figure}[t]\centering

	\centering
	\begin{tikzpicture}
	[level 1/.style={level distance=0mm},
	level 2/.style={level distance=1.5cm}]
	\coordinate
	child {node [geo] {$E_{1}$}
		edge from parent[draw=none] }
	child {node [geo,color=white,text=black] {$\cdots$}
		edge from parent[draw=none] }
	child {node [geo] {$E_{m}$}
		edge from parent[draw=none] }
	child {node [sph] {$D$}
		child {node [geo] {$\tilde{E}_{1}$}}
		child {node [geo,color=white,text=black] {$\cdots$}
			edge from parent[draw=none] }
		child {node [geo] {$\tilde{E}_{k}$}}
		edge from parent[draw=none] };
	\end{tikzpicture}
\caption{KEM web I.} \label{fig:KEMwebI}
\end{figure}

Now take $\tilde{K}$ to be the KBDT associated with a Benenti tensor on $F$ which has eigenspaces $\tilde{E}_1,\dots,\tilde{E}_k$. Then Fig.~\ref{fig:KEMwebI} is a diagram for the above construction applied to $\tilde{K}$, which represents the tree-like structure of the constructed separable web. It should be interpreted as a tree diagram, where the one dimensional eigenspaces are the leaves. We illustrate this construction with two simple examples in $\mathbb{E}^3_1$, both of which are depicted by Fig.~\ref{fig:KEMwebI} with $m = 1$ and $k = 2$.

\begin{Example}[cylindrical coordinates in $\mathbb{E}^3_1$] \label{introex:cylCoord}
	Fix a non-zero vector $d \in \mathbb{E}^3_1$ and consider the following CT:
	\begin{gather*}
	L = d \odot d.
	\end{gather*}
	
First note that if $\langle d,d\rangle = 0$, then $L$ is not diagonalizable, and hence this case can be neglected. First assume that $\langle d,d\rangle = -1$, i.e., $d$ is timelike. The eigenspaces of $L$ are then~$\operatorname{span} \{d\}$ and~$d^\perp$. Identify $\mathbb{E}_1 = \operatorname{span} \{d\}$ and $\mathbb{E}^2 = d^\perp$, then the warped product $\psi \colon \mathbb{E}_1 \times_1 \mathbb{E}^2 \rightarrow \mathbb{E}^3_1$ given by~$(q,p) \rightarrow q + p$ is adapted to the eigenspaces of~$L$. We can construct separable coordinates in~$\mathbb{E}^3_1$ by parameterizing the $\mathbb{E}^2$ factor with any of the separable coordinates in~$\mathbb{E}^2$. For example, let~$e$,~$f$ be an orthonormal basis for~$d^\perp$, let $q = t d$ and $p = \rho \cos \theta e + \rho \sin \theta f$, then we obtain cylindrical coordinates:
	\begin{gather*}
	\psi(q,p) = t d + \rho \cos \theta e + \rho \sin \theta f.
	\end{gather*}
	
	There remains the case that $d$ is spacelike, that is when $\langle d,d\rangle = 1$, then the warped product becomes $\psi \colon \mathbb{E} \times_1 \mathbb{E}^2_1 \rightarrow \mathbb{E}^3_1$, and hence separable coordinates in this space can be obtained by taking any of the separable coordinates on~$\mathbb{E}^2_1$ enumerated in Section~\ref{sec:sepE21}.
\end{Example}

The following is a more interesting example of this construction.

\begin{Example}[spherical coordinates in $\mathbb{E}^3_1$] \label{introex:sphCoord}
Consider the following CT in $\mathbb{E}^3_1$:
\begin{gather*}
	L = r \odot r.
\end{gather*}
	
The eigenspaces of $L$ are $\operatorname{span} \{r\}$ and $r^\perp$. Fix a unit vector $a \in \mathbb{E}^3_1$ with $\varepsilon = \langle a,a\rangle$, identify $\mathbb{E}_\varepsilon = \mathbb{R}^+ a$, let $\mathbb{E}^3_1(\varepsilon)$ be the unit sphere in $\mathbb{E}^3_1$ and $\rho_1 := \langle q,a\rangle$ for $q \in \mathbb{E}_\varepsilon$. Then the map $\psi \colon \mathbb{E}_\varepsilon \times_{\rho_1} \mathbb{E}^3_1(\varepsilon) \rightarrow \mathbb{E}^3_1$ given by $(q,p) \rightarrow \rho_1 p$ is a warped product adapted to the eigenspaces of~$L$. We can construct separable coordinates in $\mathbb{E}^3_1$ by parameterizing $\mathbb{E}^3_1(\varepsilon)$ with any of the separable coordinates def\/ined in it.
	
For example if $\varepsilon = 1$, then $\mathbb{E}^3_1(\varepsilon) \simeq \operatorname{dS}_2$, and one can take any of the separable coordinates in~$\operatorname{dS}_2$ enumerated in Section~\ref{sec:sepdS2}. Indeed, f\/ix a timelike unit vector $d \in a^\perp$. Then one can show that the restriction of $d \odot d$ to $\operatorname{dS}_2$ is a Benenti tensor diagonalized in spherical coordinates (see \cite[Example~9.5.13]{Rajaratnam2014}), which are given as follows
	\begin{gather*}
	p = \sinh(u) d + \cosh(u) (\sin(v) a + \cos(v) b),
	\end{gather*}
where $\{a,b,d\}$ is any orthonormal basis for $\mathbb{E}^3_1$ extending $\{a,d\}$. Hence the above coordinates are separable in $\operatorname{dS}_2$. If we let $q = \rho a$ where $\rho > 0$ and take $p$ as above, then we obtain spherical coordinates in $\mathbb{E}^3_1$:
	\begin{gather*}
	\psi(q,p) = \rho(\sinh(u) d + \cosh(u) (\sin(v) a + \cos(v) b)) ,\qquad -\infty < u < \infty,\qquad 0 < v < 2\pi.
	\end{gather*}
\end{Example}

For more details on the above example and for more general theorems on obtaining warped products decomposing CTs, see \cite[Section~9.5]{Rajaratnam2014}.

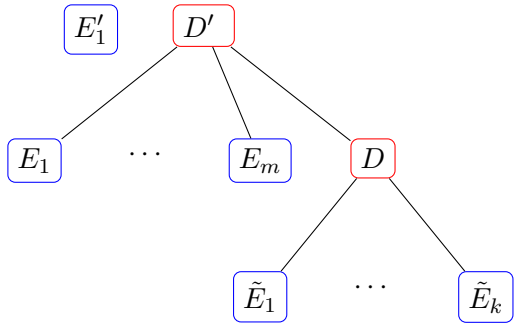
\begin{figure}[h]	\centering
	\begin{tikzpicture}
	[level 1/.style={level distance=0mm},
	level 2/.style={level distance=1.5cm}]
	\coordinate
	child {node [geo] {$E_1'$}
		edge from parent[draw=none] }
	child {node [sph] {$D'$ }
		child {node [geo] {$E_{1}$} }
		child {node [geo,color=white,text=black] {$\cdots$} edge from parent[draw=none] }
		child {node [geo] {$E_{m}$}}
		child {node [sph] {$D$}
			child {node [geo] {$\tilde{E}_{1}$}}
			child {node [geo,color=white,text=black] {$\cdots$}
				edge from parent[draw=none] }
			child {node [geo] {$\tilde{E}_{k}$}} } edge from parent[draw=none] };
	\end{tikzpicture}
	\caption{KEM web II.} \label{fig:KEMwebII}
\end{figure}

This construction procedure can be generalized in two ways. Firstly, we can recursively apply this procedure, by treating $B \times_{\rho} F$ as the spherical factor of a larger warped product and use $K + \tilde{K}$ in place of $\tilde{K}$. Fig.~\ref{fig:KEMwebII} depicts such a construction where the CT $L$ has eigenspaces $E_1'$ and $D'$. Again, this f\/igure depicts the tree-like structure of the KEM web where the leaves are the one-dimensional eigenspaces of the CTs that make it up.

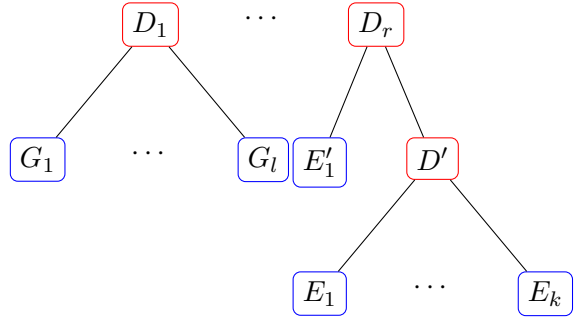
\begin{figure}[h]\centering
	\begin{tikzpicture}
	[level 1/.style={level distance=0mm},
	level 2/.style={level distance=1.5cm}]
	\coordinate
	child {node [sph] {$D_{1}$}
		child {node [geo] {$G_{1}$}}
		child {node [geo,color=white,text=black] {$\cdots$}
			edge from parent[draw=none] }
		child {node [geo] {$G_{l}$}}
		edge from parent[draw=none] }
	child {node [geo,color=white,text=black] {$\cdots$}
		edge from parent[draw=none] }
	child {node [sph] {$D_{r}$}
		child {node [geo] {$E_1'$}}
		child {node [sph] {$D'$}
			child {node [geo] {$E_{1}$}}
			child {node [geo,color=white,text=black] {$\cdots$}
				edge from parent[draw=none] }
			child {node [geo] {$E_{k}$}}
		}
		edge from parent[draw=none] };
	\end{tikzpicture}	\caption{KEM Web III.} \label{fig:KEMwebIII}
\end{figure}

Secondly, we can allow $L$ to have multiple distinct multidimensional eigenspaces. These procedures can also be combined to create even more complex webs, as the following example will show. Fig.~\ref{fig:KEMwebIII} depicts the natural generalization of the above construction procedure to CTs with multiple multidimensional eigenspaces. In this case, the CT $L$ has only multidimensional eigenspaces $D_{1},\dots,D_{r}$.

We emphasize here that in each case, the constructed web is separable. Any coordinates constructed using this procedure are called \emph{Kalnins--Eisenhart--Miller $($KEM$)$ coordinates} and the associated webs are called KEM webs. It can be shown that KEM webs are always separable \cite[Proposition~6.8]{Rajaratnam2014a}, because a ChKT can be constructed using concircular tensors as in the f\/irst example.

We've shown how CTs can be used to construct a special class of separable webs called KEM webs. A signif\/icant advantage of KEM webs is that we can reduce the problem of classifying isometrically inequivalent KEM webs to a similar problem for CTs. We've also reduced the problem of classifying isometrically inequivalent CTs in spaces of constant curvature to concrete problems in linear algebra (see Theorem~\ref{thm:conTenCanForm} for the pseudo-Euclidean case and \cite[Chapter~9]{Rajaratnam2014} for the case of spaces with non-zero constant curvature).

In conclusion, we mention how some of the ideas presented here are generalized. The observation that CTs (which are in fact CKTs) induce a warped product decomposition of the (pseudo-)Riemannian manifold, motivates the more systematic study of CKTs in~\cite{Rajaratnam2014a}. This culminates in \cite[Corollary~3.5]{Rajaratnam2014a} and \cite[Corollary~3.7]{Rajaratnam2014a}.

\subsubsection[Separable coordinates in $\operatorname{dS}_2$]{Separable coordinates in $\boldsymbol{\operatorname{dS}_2}$} \label{sec:sepdS2}

We now obtain the separable coordinates in 2-dimensional de Sitter space $\operatorname{dS}_2 = \mathbb{E}^3_1(1)$, which we realize by its standard embedding in~$\mathbb{E}^3_1$. The general concircular tensor,~$L$, on $\mathbb{E}^n_\nu(\kappa)$ is obtained by restricting the general CT on $\mathbb{E}^n_\nu$ given by equation~\eqref{eq:CTGenEunn}. Indeed, if we denote the orthogonal projection onto the spherical distribution, $r^\perp$, by~$R$, then $R$ takes the form:
\begin{gather*}
R = I - \frac{r \otimes r^{\flat}}{r^{2}},\qquad R^* = I - \frac{r^{\flat} \otimes r}{r^{2}}.
\end{gather*}

Then the general CT, $L$, in $\mathbb{E}^{n}_{\nu}(\kappa)$ depends only on a constant two-tensor $A = A^{ij} \partial_i \odot \partial_j$ in~$\mathbb{E}^{n}_{\nu}$, and is given as follows in contravariant form \cite[Proposition~9.3.2]{Rajaratnam2014}:
\begin{gather*} 
L = R A R^* = A + \kappa^2 \langle r, A r\rangle r \odot r - 2 \kappa (A r \odot r), \qquad L^{ij} = R^i_{~l} A^{lk} R^j_{~k}
\end{gather*}
in the ambient pseudo-Euclidean space.

As in the case of $\mathbb{E}^2_1$, since $\operatorname{dS}_2$ is two dimensional, we need only classify the non-trivial Benenti tensors in $\operatorname{dS}_2$. By the above equation for $L$, the problem of classifying the inequivalent Benenti tensors on $\operatorname{dS}_2$ then reduces to the classif\/ication of certain canonical forms for $A$ (these canonical forms are essentially the metric-Jordan canonical form of the pair $(A, g)$, see, e.g., \cite[Section~9.3.2]{Rajaratnam2014} for details). For the sake of brevity, we outline the procedure for two particular cases, and summarize the f\/inal results in the subsequent table. As usual, the details of the theory shall be left to the interested reader. The classif\/ication is similar to the one in $\mathbb{E}^2_1$, where now we use formulas for the characteristic polynomial of~$L$ from \cite[Section~9.4.3]{Rajaratnam2014}. Throughout this section, we let $(t,x,y)$ denote the ambient Cartesian coordinates in~$\mathbb{E}^3_1$.

{\bf Case 1, $\boldsymbol{A = J_{-1}(\lambda_1) \oplus J_1(\lambda_2) \oplus J_1(\lambda_3)}$ and $\boldsymbol{\lambda_1 < \lambda_2 < \lambda_3}$.} In this case, the induced CT is irreducible and the equations for the canonical coordinates $(u, v)$ are readily obtained. By geometric equivalence, we let $(\lambda_1, \lambda_3) \rightarrow (1, 0)$ and rename $\lambda_2 = a^2$. Now letting $(u, v) \rightarrow (f_1(u), f_2(v))$ and requiring the metric to be in Liouville form, we obtain the following transformation equations in terms of the Jacobi elliptic functions:
\begin{gather*}t^2 = \operatorname{sc}^2(u;a) \operatorname{dn}^2(v;a),\qquad x^2 = \operatorname{nc}^2(u;a) \operatorname{cn}^2(v;a),\qquad y^2 = \operatorname{dc}^2(u;a) \operatorname{sn}^2(v;a),\end{gather*}
where $0 < v,u < K(a)$ and $K(a)$ is the complete elliptic integral of the f\/irst kind with parameter~$a$. The metric then takes the form
\begin{gather*} ds^2 = \big(\operatorname{dc}^2(u;a) - a^2 \operatorname{sn}^2(v;a)\big)\big({-} du^2 + dv^2\big).\end{gather*}

{\bf Case 3, $\boldsymbol{A = J_{-1}(\lambda_1) \oplus J_1(\lambda_2) \oplus J_1(\lambda_3)}$ and $\boldsymbol{\lambda_1 > \lambda_2 = \lambda_3}$.} By geometric equivalence, we may let $(\lambda_1, \lambda_2, \lambda_3) \rightarrow (1,0,0)$. In this case, one can show the induced CT is reducible. We construct a warped product which decomposes $L$, given by $\psi \colon \operatorname{dS}_1 \times_\rho \mathbb{S}^1 \rightarrow \operatorname{dS}_2$. We identify~$\operatorname{dS}_1$ with the unit de Sitter circle in the $t$-$x$ plane, and $\mathbb{S}^1$ with the unit circle in the $x$-$y$ plane. Upon choosing the standard coordinates on each of the factors, we have, for $-\infty < u < \infty$ and $0 < v < 2\pi$,
\begin{gather*}
t = \sinh{u},\qquad x = \cosh{u} \cos{v},\qquad y = \cosh{u} \sin{v},\qquad ds^2 = -du^2 + \cosh^2{u} dv^2.
\end{gather*}
One can continue in this manner, classifying the inequivalent webs on $\operatorname{dS}_2$ according to their inducing constant two-tensor $A$, and obtain the metric and transformation equations as demonstrated. We tabulate the result of this classif\/ication in Table~\ref{tab:dS2} below. We give the metrics, transformation equations and coordinate ranges, as well as (a canonical choice of) the corresponding~$A$. Further note that while we often give the equations for a single chart domain, the corresponding equations for all other domains covering the web may be obtained by isometry (for instance, some combination of $t \rightarrow \pm t$, $x \rightarrow \pm x$ and $y \rightarrow \pm y$).

\begin{center}
	\begin{longtable}{|l|l|}
		\caption{Separable Coordinate Systems in dS$_2$.} \label{tab:dS2} \\
		\hline
		\textbf{Case 1:} 							& $ds^2 = (\operatorname{dc}^2(u;a) - a^2 \operatorname{sn}^2(v;a))(-du^2 + dv^2)$ \tsep{2pt}\\
		Real elliptic coordinates of Type I 				& $t = \operatorname{sc}(u;a) \operatorname{dn}(v;a)$ \\
		$A = J_{-1}(0) \oplus J_1(a^2) \oplus J_1(1)$ 	 & $x= \operatorname{nc}(u;a) \operatorname{cn}(v;a)$ \\
		$0 < a < 1$ 								 & $y = \operatorname{dc}(u;a) \operatorname{sn}(v;a)$ \\
		& $0 < u < K(a)$, $0 < v < K(a)$ \\
		\hline
		\textbf{Case 2:} 							& for $a|t| - |x| > b$ \\
		Real elliptic coordinates of Type II				& $ds^2 = (\operatorname{dc}^2(u;a) - \operatorname{dc}^2(v;a))(-du^2 + dv^2)$ \\
		$A = J_{-1}(a^2) \oplus J_1(0) \oplus J_1(1)$ 	& $t = \frac{b}{a} \operatorname{nc}(u;a) \operatorname{nc}(v;a)$ \\
		$0 < a < 1$ 								& $x= b \operatorname{sc}(u;a) \operatorname{sc}(v;a)$ \\
		& $y = \frac{1}{a} \operatorname{dc}(u;a) \operatorname{dc}(v;a)$ \\
		& $0 < v < u < K(a)$, $a^2 + b^2 = 1$\bsep{3pt} \\
		& for $a|t| + |x| < b$ \\
		& $ds^2 = a^2(\operatorname{nd}^2(u;b) - \operatorname{nd}^2(v;b))(du^2 - dv^2)$ \\
		& $t = a b \operatorname{sd}(u;b) \operatorname{sd}(v;b)$ \\
		& $x = b \operatorname{cd}(u;b) \operatorname{cd}(v;b)$ \\
		& $y = a \operatorname{nd}(u;b) \operatorname{nd}(v;b)$ \\
		& $0 < v < u < K(b)$, $a^2 + b^2 = 1$ \\
		\hline
		\textbf{Case 3:} 							& $ds^2 = -du^2 + \cosh^2{u} dv^2$\tsep{2pt} \\
		Spherical coordinates of Type I 				& $t = \sinh{u}$ \\
		$A = J_{-1}(1) \oplus J_1(0) \oplus J_1(0)$ 	& $x= \cosh{u} \cos{v}$ \\
		& $y = \cosh{u} \sin{v}$ \\
		& $-\infty < u < \infty$, $0 < v < 2\pi$ \\
		\hline
		\textbf{Case 4:} 							& for $-t^2 + x^2 > 0$ \tsep{2pt}\\
		Spherical Coordinates of Type II				& $ds^2 = du^2 - \sin^2{u} dv^2$ \\
		$A = J_{-1}(0) \oplus J_1(0) \oplus J_1(1)$ 	& $t = \sin{u} \sinh{v}$ \\
		& $x= \sin{u} \cosh{v}$ \\
		& $y = \cos{u}$ \\
		& $0 < u < \pi$, $-\infty < v < \infty$ \bsep{3pt}\\
		& for $-t^2 + x^2 < 0$ \\
		& $ds^2 = -du^2 + \sinh^2{u} dv^2$ \\
		& $t = \sinh{u} \cosh{v}$ \\
		& $x= \sinh{u} \sinh{v}$ \\
		& $y = \cosh{u}$ \\
		& $0 < u < \infty$, $-\infty < v < \infty$ \\
		\hline
		\textbf{Case 5:} 									& $ds^2 = f(u,v)(-du^2 + dv^2)$\tsep{2pt} \\
		Complex elliptic coordinates							& $f(u,v) = \operatorname{sn}^2(u;a)\operatorname{dc}^2(u;a) - \operatorname{sn}^2(v;a)\operatorname{dc}^2(v;a)$ \\
		$A = J_{1}(i \beta) \oplus J_1(- i \beta) \oplus J_1(c)$	& $t^2 + x^2 = \dfrac{2\operatorname{dn}(2u;a) \operatorname{dn}(2v;a)}{ab(1+\operatorname{cn}(2u;a))(1+\operatorname{cn}(2v;a))}$ \\
		$\beta > 0$									 	& $-t^2 + x^2 = \dfrac{2(\operatorname{cn}(2u;a) + \operatorname{cn}(2v;a))}{(1 + \operatorname{cn}(2u;a))(1 + \operatorname{cn}(2v;a))}$ \\
		& $y = \operatorname{sn}(u;a) \operatorname{dc}(u;a)\operatorname{sn}(v;a) \operatorname{dc}(v;a)$ \\
		& $0 < v < u < K(a)$, $a^2 + b^2 = 1$ \\
		\hline
		\textbf{Case 6:} 									& $ds^2 = (\operatorname{sech}^2{u} + \operatorname{csch}^2{v})(du^2 - dv^2)$ \tsep{2pt}\\
		Null elliptic coordinates of Type I						& $t + x = \operatorname{sech}{u} \operatorname{csch}{v}$ \\
		$A = J_2(0)^T \oplus J_1(c)$						& $t - x = - \cosh{u} \sinh{v} (1 - \tanh^2{u} \coth^2{v})$ \\
		$c > 0$										 	& $y = \tanh{u} \coth{v}$ \\
		& $0 < u < \infty$, $0 < v < \infty$ \\
		\hline
		\textbf{Case 7:} 							& for $|x| > 1$, $tx > 0$ \\
		Null elliptic coordinates of Type II				& $ds^2 = (\sec^2{u} - \sec^2{v})(-du^2 + dv^2)$ \\
		$A = J_{-2}(0)^T \oplus J_1(c)$			 	& $t + x = \sec{u} \sec{v}$ \\
		$c > 0$									& $t - x = - \cos{u} \cos{v} (1 - \tan^2{u} \tan^2{v})$ \\
		& $y = \tan{u} \tan{v}$ \\
		& $0 < v < u < \frac{\pi}{2}$ \bsep{3pt}\\
		&for $|x| > 1$, $tx < 0$, $|y| > 1$ \\
		& $ds^2 = (\operatorname{csch}^2{v} - \operatorname{csch}^2{u})(du^2 - dv^2)$ \\
		& $t + x = \operatorname{csch}{u} \operatorname{csch}{v}$ \\
		& $t - x = - \sinh{u} \sinh{v} (1 - \coth^2{u} \coth^2{v})$ \\
		& $y = \coth{u} \coth{v}$ \\
		& $0 < v < u < \infty$ \bsep{3pt}\\
		& for $|x| > 1$, $tx < 0$, $|y| < 1$ \\
		& $ds^2 = (\operatorname{sech}^2{u} - \operatorname{sech}^2{v})(du^2 - dv^2)$ \\
		& $t + x = \operatorname{sech}{u} \operatorname{sech}{v}$ \\
		& $t - x = - \cosh{u} \cosh{v} (1 - \tanh^2{u} \tanh^2{v})$ \\
		& $y = \tanh{u} \tanh{v}$ \\
		& $0 < u < v < \infty$ \\
		\hline
		\textbf{Case 8:} 									& $ds^2 = -du^2 + e^{2u}dv^2$ \tsep{2pt}\\
		Null spherical coordinates							& $t+x = e^{-u} - v^2e^u$ \\
		$A = J_2(0)^T \oplus J_1(0)$						& $t-x = -e^u$ \\
		& $y = ve^u$ \\
		& $-\infty < u < \infty$, $-\infty < v < \infty$ \\
		\hline
		\textbf{Case 9:} 									& $ds^2 = (u^{-2} - v^{-2})(-du^2 + dv^2)$ \tsep{2pt}\\
		Null elliptic coordinates of Type III					& $t+x = \dfrac{1}{uv}$ \\
		$A = J_3(0)^T$									& $t-x = \dfrac{(u^2 - v^2)^2}{4uv}$ \\
		& $y = \dfrac{u^2 + v^2}{2uv}$ \\
		& $0 < u < v < \infty$ \\
		\hline
	\end{longtable}
\end{center}

As in the classif\/ication of separable coordinates in $\mathbb{E}^2_1$, this classif\/ication is exhaustive due to the KEM separation theorem \cite[Theorem~1.4]{Rajaratnam2014d}, which when applied to $dS_{2}$ says: in $\operatorname{dS}_2$ any separable coordinate system admits a non-trivial Benenti tensor which is diagonalized in the coordinates. See the remarks at the end of Section~\ref{sec:sepE21} to see how one would carry out this classif\/ication in higher dimensions and signatures.

Additionally, one may readily obtain the separable coordinates on $\operatorname{AdS}_2$, $2$-dimensional anti-de Sitter space, by the following prescription: if $ds^2 = -g_{11}(u,v)du^2 + g_{22}(u,v) dv^2$ is a~metric for~$\operatorname{dS}_2$, then the metric obtained by letting $g_{11}(u,v) \rightarrow g_{22}(v,u)$ and $g_{22}(u,v) \rightarrow g_{11}(v,u)$ is a~metric on~$\operatorname{AdS}_2$. Moreover, the transformation formulae are obtained by simultaneously letting $(t,y) \rightarrow (y,t)$ and $(u,v) \rightarrow (v,u)$ in the equations above. This correspondence occurs because $\operatorname{AdS}_2$ can be embedded in~$\mathbb{E}^3_2$, which is simply $\mathbb{E}^3_1$ with a reversal of signature. Separable webs in 2-dimensional spaces of constant curvature are also considered in~\cite{Carinena2005,Horwood2008a}. However, it seems that these treatments are not exhaustive.

\begin{figure}[t]	\centering
	\begin{subfigure}{0.3\textwidth}
		\includegraphics[scale=0.3]{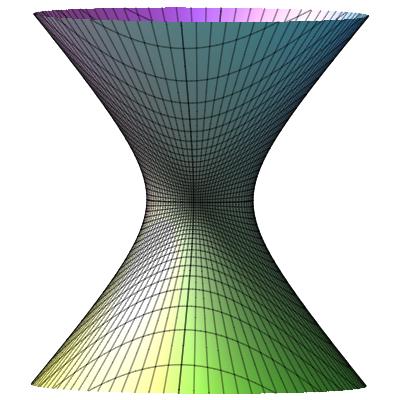}
		\caption{Real elliptic I}		\label{fig:dS1}
	\end{subfigure}
	\begin{subfigure}{0.3\textwidth}
		\includegraphics[scale=0.3]{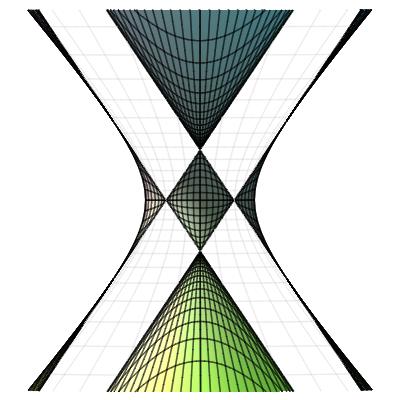}
		\caption{Real elliptic II}
		\label{fig:dS2}
	\end{subfigure}
	\begin{subfigure}{0.3\textwidth}
		\includegraphics[scale=0.3]{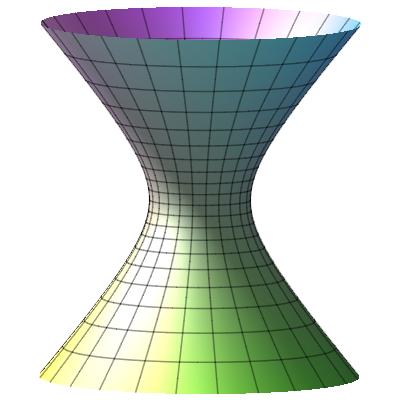}
		\caption{Spherical I}
		\label{fig:dS3}
	\end{subfigure}
	\\
	\begin{subfigure}{0.3\textwidth}
		\includegraphics[scale=0.3]{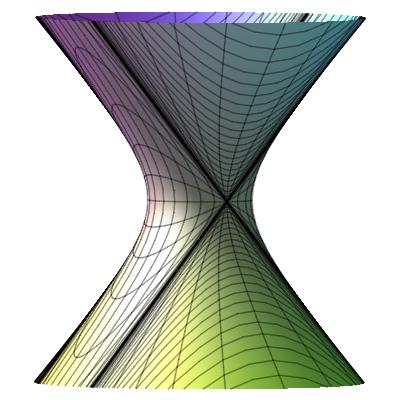}
		\caption{Spherical II}
		\label{fig:dS4}
	\end{subfigure}
	\begin{subfigure}{0.3\textwidth}
		\includegraphics[scale=0.3]{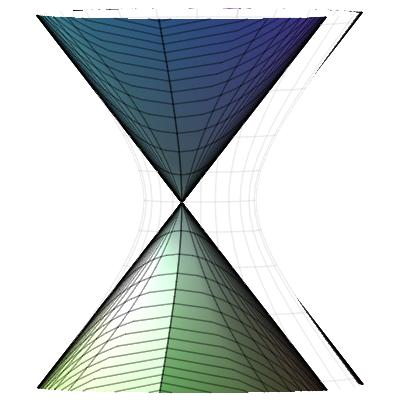}
		\caption{Complex elliptic}
		\label{fig:dS5}
	\end{subfigure}
	\begin{subfigure}{0.3\textwidth}
		\includegraphics[scale=0.3]{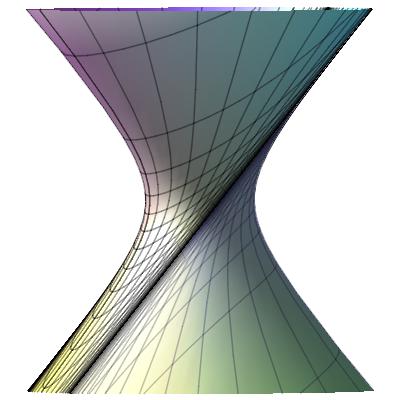}
		\caption{Null elliptic I}
		\label{fig:dS6}
	\end{subfigure}
	\\
	\begin{subfigure}{0.3\textwidth}
		\includegraphics[scale=0.3]{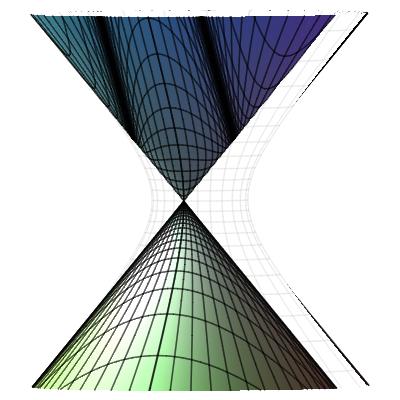}
		\caption{Null elliptic II}
		\label{fig:dS7}
	\end{subfigure}
	\begin{subfigure}{0.3\textwidth}
		\includegraphics[scale=0.3]{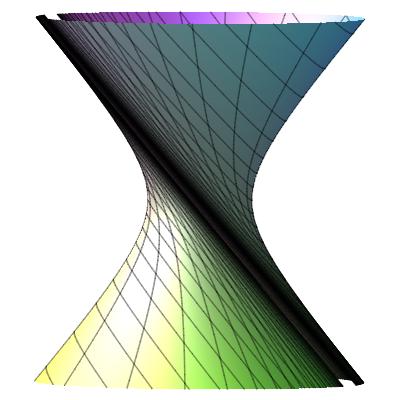}
		\caption{Null spherical}
		\label{fig:dS8}
	\end{subfigure}
	\begin{subfigure}{0.3\textwidth}
		\includegraphics[scale=0.3]{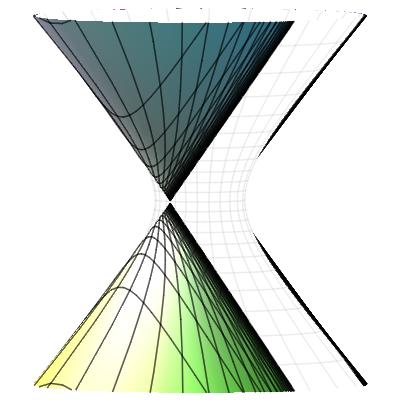}
		\caption{Null elliptic III}
		\label{fig:dS9}
	\end{subfigure}
		\caption{Separable coordinate webs on $\operatorname{dS}_2$.}
	\label{fig:dS2webs}
\end{figure}

On the next page, we provide some illustrations of the separable webs on $\operatorname{dS}_2$ (Fig.~\ref{fig:dS2webs}), obtained using Maple; we employ the standard representation of $\operatorname{dS}_2$ as a hyperboloid of one sheet embedded in $\mathbb{E}^3_1$. The reader should compare these illustrations with those of the webs on~$\mathbb{E}^2_1$ (see Fig.~\ref{fig:M2webs}) and note the similarities. In Fig.~\ref{fig:dS2webs}, as in Fig.~\ref{fig:M2webs}, the empty white spaces containing no coordinate curves represent the open singular sets of the webs, and the dark lines represent the closed singular sets, often separating dif\/ferent regions of the web. Finally, just as in~$\mathbb{E}^2_1$, we note that the existence of these open singular sets, as well as the existence of inequivalent\footnote{Here, by inequivalent domains we mean domains whose webs cannot be mapped into each other by isometry, e.g., Figs.~\ref{fig:dS2},~\ref{fig:dS4} and~\ref{fig:dS7}.} coordinate domains, ref\/lect how the introduction of a Lorentzian signature gives rise to a much richer theory of separable coordinates than in the Riemannian case.

\subsection{Necessity of KEM webs in spaces of constant curvature} \label{sec:KEMsep}

In the previous section we have shown how to construct a class of separable webs called KEM webs. These webs were originally discovered by Kalnins and Miller when classifying the separable webs in Riemannian spaces of constant curvature~\cite{Kalnins1986}. Generalizing their results, one can prove the following.

\begin{Theorem}[separable webs in spaces of constant curvature \cite{Rajaratnam2014d}] \label{Intthm:SCCconChKT}
	In a space of constant curvature, every orthogonal separable web is a KEM web.
\end{Theorem}

This theorem allows us to tractably solve Problem~(\ref{fpro:I}) in spaces of constant curvature. It proves that the classif\/ication of separable coordinates in Section~\ref{sec:sepE21} is complete, and gives a tractable method to enumerate the separable coordinates in $\mathbb{E}^n_1$ (see the remarks at the end of Section~\ref{sec:sepE21}) and more generally, all spaces of constant curvature.

The above theorem is a consequence of the following one.

\begin{Theorem}[KEM separation theorem \cite{Rajaratnam2014d}] \label{thm:KEMsep}
	Suppose $K$ is a ChKT defined on a space of constant curvature $M$. Then there is a non-trivial concircular tensor $L$ defined on~$M$ such that each eigenspace of $K$ is L-invariant, i.e., $L$ is diagonalized in coordinates adapted to the eigenspaces of~$K$.
\end{Theorem}

Let $K$ be an arbitrary ChKT def\/ined on a space of constant curvature $M$. Then the above theorem guarantees the existence of a non-trivial CT,~$L$, which algebraically commutes with~$K$. If~$L$ has simple eigenfunctions, then $L$ is a Benenti tensor and induces a separable web (see Section~\ref{sec:geo}), with $K$ in the associated KS-space.

Otherwise, $K$ has multidimensional eigenspaces. For simplicity we assume $K$ has a single multidimensional eigenspace~$D$. Then as in Section~\ref{sec:CTsMultEig}, the pair $(D^\perp, D)$ induces a warped product $B \times_\rho F$ which is locally isometric to~$M$. One can show that both~$B$ and~$F$ are spaces of constant curvature with $B$ having the same curvature as~$M$ \cite[Section~4]{Rajaratnam2014d}.Then by Proposition~\ref{prop:wpKss}, $K$ induces a Killing tensor~$K_F$ on~$F$, by restriction. One can then recursively construct the KEM web by applying the above theorem to the ChKT $K_F$ on the space of constant curvature $F$, and then analyzing the resulting concircular tensor as above. See \cite[Section~6.2]{Rajaratnam2014a} for details.

As the above discussion shows, the above theorem is the key to classifying all separable webs in spaces of constant curvature. However, the proof of this theorem, given in~\cite{Rajaratnam2014d}, involves a long calculation in which we solve the Levi-Civita equations together with the equations satisf\/ied by the Riemann curvature tensor in a space of constant curvature. An important property of KEM webs that we use in this classif\/ication is that they have \emph{diagonal curvature} \cite[Proposition~6.5.5]{Rajaratnam2014}. Indeed one can show that in any KEM coordinate system~$(x^i)$, the Riemann curvature tensor satisf\/ies $R_{ijik} = 0$ for $j \neq k$, which is called the \emph{diagonal curvature condition}. This condition is equivalent to requiring the curvature operator (which is a~$\binom{2}{2}$-tensor associated with~$R$ which induces a map in $\operatorname{End}(\wedge^{2}(M))$~\cite{Peter2006}) to be diagonal in the coordinate induced basis.

\section{Separation of natural Hamiltonians} \label{sec:sepNats}

In this section we will sketch how concircular tensors can be used to separate natural Hamiltonians. We will use Theorem~\ref{Intthm:HJosepII} and our knowledge of the structure of KEM webs to develop a~recursive algorithm to separate natural Hamiltonians in KEM webs.

Fix some $V \in \mathcal{F}(M)$ and assume $n \geq 2$ to avoid trivial cases. Let $L$ be the general concircular tensor on $M$ and let $K := \operatorname{tr}(L)G - L$ be the KBDT associated with~$L$. The Killing--Bertrand--Darboux (KBD) equation on $M$ is def\/ined as follows
\begin{gather*}
{\rm d} (K {\rm d} V) = 0.
\end{gather*}

It can be shown that this equation def\/ines a linear system of equations with at most $\frac{1}{2}(n+1)(n+2)$ unknowns, where the maximum is achieved if\/f the space has constant curvature.

Let $L$ be a particular solution of the KBD equation which is point-wise diagonalizable with~$k$ distinct eigenfunctions. We analyze the following cases.

{\bf Case 1} ($k = 1$, i.e., all the eigenfunctions coincide).
	$L = c G$ for some $c \in \mathbb{R}$. This is the trivial solution which gives no information.

{\bf Case 2} (the eigenfunctions are simple).
	$L$ has simple eigenfunctions, hence it's a Benenti tensor. Then $V$ separates in any coordinates which diagonalize $L$ by Theorem~\ref{Intthm:HJosepII}.

\begin{figure}[h]		\centering
		\begin{tikzpicture}
		[level 1/.style={level distance=0mm},
		level 2/.style={level distance=1.5cm}]
		\coordinate
		child {node [geo] {$E_{1}$}
			edge from parent[draw=none] }
		child {node [geo,color=white,text=black] {$\cdots$}
			edge from parent[draw=none] }
		child {node [geo] {$E_{m}$}
			edge from parent[draw=none] }
		child {node [sph] {$D$}
			edge from parent[draw=none] };
		\end{tikzpicture}
\caption{Concircular tensor with eigenspaces $E_{1},\dots,E_{m},D$.} \label{fig:KEMwebPart}
	\end{figure}
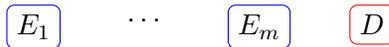

{\bf Case 3} (at least one eigenfunction is not simple).
	Assume for convenience, that~$L$ has a~single multidimensional eigenspace~$D$. If $E_1,\dots,E_m$ denote the one dimensional eigenspaces of~$L$, then so far we know that $V$ is ``compatible'' with the partial separable web in Fig.~\ref{fig:KEMwebPart}.
	
	Now the goal is to f\/ill in the degeneracy coming from $D$. This is done as follows: let $B \times_{\rho} F$ be a local warped product adapted to $(D^{\perp}, D)$. Let $\tau\colon F \rightarrow B \times F$ be an embedding. Assume the natural Hamiltonian on $F$ associated with potential $V \circ \tau$ is separable in some coordinates~$(y^j)$. Let~$(x^i)$ be separable coordinates associated with the induced Benenti tensor on~$B$. Then one can show that the natural Hamiltonian associated with $V$ (on $B \times_{\rho} F$) is separable in the product coordinates~$(x^i,y^j)$.
	
	Indeed, this can be seen as follows: let $\tilde{K}$ be a ChKT on $F$ diagonalized in $(y^j)$, and $K'$ be the KBDT associated with $L$. In the discussion preceding equation~\eqref{introeq:KTsumKEM}, it was shown that we can assume $K := K' + \tilde{K}$ is locally a ChKT on $B \times_{\rho} F$ diagonalized in $(x^i,y^j)$. Given the assumptions, one can show that~$V$ satisf\/ies the dKdV equation with $K$ on $B \times_{\rho} F$, hence by Theorem~\ref{Intthm:HJosepII} it's separable in the coordinates $(x^i,y^j)$.

\begin{figure}[h]	\centering
	\begin{tikzpicture}
	[level 1/.style={level distance=0mm},
	level 2/.style={level distance=1.5cm}]
	\coordinate
	child {node [geo] {$E_{1}$}
		edge from parent[draw=none] }
	child {node [geo,color=white,text=black] {$\cdots$}
		edge from parent[draw=none] }
	child {node [geo] {$E_{m}$}
		edge from parent[draw=none] }
	child {node [sph] {$D$}
		child {node [geo] {$\tilde{E}_{1}$}}
		child {node [geo,color=white,text=black] {$\cdots$}
			edge from parent[draw=none] }
		child {node [geo] {$\tilde{E}_{k}$}}
		edge from parent[draw=none] };
	\end{tikzpicture}
	\caption{Possible KEM web that can be constructed.} \label{fig:KEMwebPos}
\end{figure}
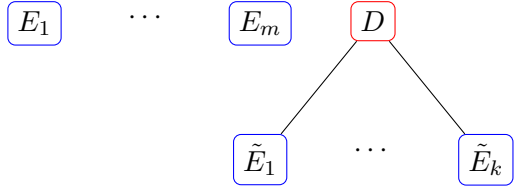

In the third case, in order to obtain the separable coordinates $(y^j)$, the idea is to apply the same procedure again on $F$ with the potential $V \circ \tau \in \mathcal{F}(F)$. So one has to solve the KBD equation on $F$ with the potential $V \circ \tau$ and then go through each case. This gives us a recursive algorithm for separating natural Hamiltonians, which is called the Benenti--Eisenhart--Kalnins--Miller (BEKM) separation algorithm. This algorithm is presented in more detail and with proofs in \cite[Section~6.3]{Rajaratnam2014a}. Fig.~\ref{fig:KEMwebPos} gives a possible KEM web that can be constructed, assuming the solution of the KBD equation on $F$ is a Benenti tensor with eigenspaces $\tilde{E}_{1},\dots,\tilde{E}_{k}$.

In principle, one can construct any KEM web using the BEKM separation algorithm. Indeed, if the BEKM separation algorithm is applied with $V = 0$, then one can construct all KEM webs in the underlying space by following through the steps of the algorithm. We now brief\/ly illustrate the execution of this algorithm with the following example.

\begin{Example}[Calogero--Moser system]
The Calogero--Moser system is a natural Hamiltonian system with conf\/iguration manifold $\mathbb{E}^3$ given by the following potential in Cartesian coordinates $(q_{1},q_{2},q_{3})$:
	\begin{gather*}
	V = (q_{1} - q_{2})^{-2} + (q_{2} - q_{3})^{-2} + (q_{1} - q_{3})^{-2}.
	\end{gather*}
	
	First note that the constant vector $d = \frac{1}{\sqrt{3}} (\partial_1 + \partial_2 + \partial_3)$ is a symmetry of $V$, i.e., $\mathcal{L}_d V = 0$. One can prove that the general solution of the KBD equation associated with $V$ is\footnote{We ignore constant multiples of the metric.}
	\begin{gather*} 
	L = c d \odot d + 2 w d \odot r + m r \odot r,
	\end{gather*}
where $c,w,m \in \mathbb{R}$. We note that given a CT $L$, then for any $a \in \mathbb{R} \setminus \{0\}$ and $b \in \mathbb{R}$, the CT $a L + b G$ is a CT which is equivalent to $L$. After classifying the above CTs modulo this equivalence and isometric equivalence, we can obtain canonical forms. Before we present these, f\/ix an orthonormal basis~$e$,~$f$ for~$d^\perp$. We have the following canonical forms.
	
{\bf Cartesian:} $L = d \odot d$. As in Example~\ref{introex:cylCoord}, one can show that a warped product manifold adapted to~$L$ has the form $\mathbb{E} \times \mathbb{E}^2$. Let $(q_1',q_2',q_3')$ be Cartesian coordinates adapted to this product manifold, then one can show that $V$ takes the form
		\begin{gather*}
		V = \frac{9\big(q_{3}'^2 + q_{2}'^2 \big)^2}{2 q_{2}'^2 \big(3 q_3'^2 - q_{2}'^2\big)^2}.
		\end{gather*}
		
		In this case $V$ naturally restricts to a potential on $\mathbb{E}^2$ with coordinates $(q_2',q_3')$. In $\mathbb{E}^2$ one can apply the BEKM separation algorithm to f\/ind that the only solution of the KBD equation (up to constant multiplies) is $L = r \odot r$. One can show that polar coordinates diagonalize this CT. Hence $V$ is separable in cylindrical coordinates
		\begin{gather*}
		x d + \rho \cos \theta e + \rho \sin \theta f.
		\end{gather*}
		
{\bf Spherical:} $L = r \odot r$.
		As in Example~\ref{introex:sphCoord}, one can show that a warped product manifold adapted to~$L$ has the form $\mathbb{E} \times_{\rho} \mathbb{S}^2$. One can show that the restriction of $V$ to $\mathbb{S}^2$ satisf\/ies the KBD equation associated with the CT obtained by restricting $d \odot d$ to $\mathbb{S}^2$. Hence from Example~\ref{introex:sphCoord}, $V$ is separable in spherical coordinates:
		\begin{gather*}
		\rho(\cos(\phi) d + \sin(\phi) (\cos(\theta) e + \sin(\theta) f)).
		\end{gather*}
		
{\bf Elliptic:} $L = c d \odot d + r \odot r$, $c \neq 0$. In this case $L$ is a Benenti tensor. If we let $a := \sqrt{|c|}$, then if $c > 0$, $V$ is separable in prolate spheroidal coordinates:
		\begin{gather*}
		a\cos \phi \cosh \eta d + a\sin \phi \sinh \eta (\cos \theta e + \sin \theta f).
		\end{gather*}
		
		If $c < 0$, $V$ is separable in oblate spheroidal coordinates:
		\begin{gather*}
		a\sin \phi \sinh \eta d + a\cos \phi \cosh \eta (\cos \theta e + \sin \theta f).
		\end{gather*}

{\bf Parabolic:} $L = 2 d \odot r$. In this case $L$ is a Benenti tensor, and so $V$ is separable in rotationally symmetric parabolic coordinates:
		\begin{gather*}
		\frac{1}{2}\big(\mu^{2} - \nu^{2}\big) d + \mu\nu (\cos \theta e + \sin \theta f).
		\end{gather*}
\end{Example}

The above example is done in much greater detail and for a more general potential in~\cite{Rajaratnam2014}. The separability properties of the above system have been studied by several dif\/ferent authors \cite{Benenti2000,Calogero1969, Horwood2005,Rauch-Wojciechowski2005,Waksjo2003}. The above solution using the BEKM separation algorithm is either comparable to these studies or more direct and concise.

The following example is a new application of the BEKM separation algorithm to a natural Hamiltonian in~$\mathbb{E}^3_1$.

\begin{Example}[Morosi--Tondo system]
Hamiltonian systems often arise naturally from the stationary f\/lows of soliton equations such as the celebrated KdV equation. In~\cite{Morosi1997}, Morosi and Tondo considered the integrability of the natural Hamiltonian system with conf\/iguration manifold $\mathbb{E}^3_1$, given by the following potential in coordinates $(\mu, \nu, y)$, where $(\mu, \nu)$ are lightlike coordinates associated with Cartesian coordinates $(t,x)$ such that $\langle \partial_\mu, \partial_\nu \rangle = 1$
\begin{gather*}V = -\frac{5}{8}\mu^4 + \frac{5}{2}\mu^2\nu + \frac{1}{2}\mu y^2 - \frac{1}{2}\nu^2,\end{gather*}
which is obtained as a stationary reduction of the seventh-order KdV f\/low. One can show that the general solution of the KBD equation associated with this $V$ is, upon ignoring constant multiples of the metric and rescaling, $L = A + 2 w \odot r$, where in coordinates $(\mu, \nu, y)$, $A = J_2(0)^T \oplus J_1(0)$ and $w = \partial_\nu$. Thus we have the canonical form for the general solution~$L$, and see that modulo geometric equivalence, only one CT is compatible with this potential. Since~$L$ is an ICT, the transformation equations from canonical coordinates $(u,v,w)$ to $(\mu, \nu, y)$ are readily obtained:
\begin{gather*}
\mu = \frac{1}{8}\big((u-v)^2+w^2\big)-\frac{1}{4}w(u+v),\qquad \nu = \frac{1}{2}(u+v+w),\qquad y^2 = uvw\end{gather*}
	with $w < v < 0 < u$. The metric in these coordinates takes the form
\begin{gather*} ds^2 = \frac{(u-v)(u-w)}{4u}du^2 - \frac{(u-v)(v-w)}{4v}dv^2 + \frac{(v-w)(u-w)}{4w}dw^2.\end{gather*}
	One may then use equations~\eqref{introeq:BTLeq} and~\eqref{introeq:BTLseqb} to obtain the following two Killing tensors which, along with the metric, span the Killing--St\"ackel space, and can be used to obtain the f\/irst integrals (we give the components of the Killing tensors in Cartesian coordinates associated with $(\mu, \nu, y)$)	
\begin{gather*} K_1^{ij} = \begin{pmatrix}
	2\sqrt{2}x - 1 & \sqrt{2}(t-x) + 1 & \sqrt{2}y \\
	\sqrt{2}(t-x) + 1 & -2\sqrt{2}t - 1 & -\sqrt{2}y \\
	\sqrt{2}y & -\sqrt{2}y & -2\sqrt{2}(t+x)
	\end{pmatrix}^{ij}, \\
K_2^{ij} = \begin{pmatrix}
	y^2 & -y^2 & -2(t+x)y \\
	-y^2 & y^2 & 2(t+x)y \\
	-2(t+x)y & 2(t+x)y & (t+x)^2
	\end{pmatrix}^{ij}.
\end{gather*}
	
	This result agrees with that obtained by Horwood, McLenaghan and Smirnov~\cite{Horwood2009} (up to isometry and linear combinations) by using the invariant theory of Killing tensors. However, the analysis presented here seems to be more direct and concise.
\end{Example}

{\bf Completeness of the BEKM separation algorithm.} In spaces of constant curvature, the BEKM separation algorithm gives a complete test for orthogonal separation. This is a~consequence of Theorem~\ref{Intthm:SCCconChKT}. We also note that the separable coordinates can be explicitly constructed by following through the algorithm, as can be seen in the above examples. Hence the BEKM separation algorithm solves Problems~(\ref{fpro:II}) and~(\ref{fpro:III}) in spaces of constant curvature.

{\bf Spaces of constant curvature.} In order to apply the BEKM separation algorithm (i.e., reduce it to problems in linear algebra) in spaces of constant curvature, CTs in these spaces are studied throughly in \cite[Chapter~9]{Rajaratnam2014}. The results from this chapter are used to apply the BEKM separation algorithm, in more detail than above, to study the separability of the Calgero--Moser system in \cite[Chapter~10]{Rajaratnam2014}.

\subsection*{Acknowledgements}

We would like to thank the referees for their helpful comments and suggestions. This work was supported in part by a QEII-Graduate Scholarship in Science and Technology (KR), Natural Sciences and Engineering Research Council of Canada Discovery Grant (RGM) and Undergraduate Student Research Award (CV).

\pdfbookmark[1]{References}{ref}
\LastPageEnding

\end{document}